\documentclass[9pt,twocolumn,twoside]{pnas-new}

\templatetype{pnasresearcharticle} 

\newcommand{\xie}{\color{black}}
\newcommand{\dong}{\color{black}}
\newcommand{\blue}{\color{blue}}

\newcommand{\beginSI}{%
        \setcounter{table}{0}
        \renewcommand{\thetable}{S\arabic{table}}%
        \setcounter{equation}{0}
        \renewcommand{\theequation}{S\arabic{equation}}%
        \setcounter{figure}{0}
        \renewcommand{\thefigure}{S\arabic{figure}}%
         }

\title{Exoplanet Orbital Eccentricities Derived From LAMOST-Kepler Analysis}

\author[a,b,1,2]{Ji-Wei Xie}
\author[c,1,2]{Subo Dong} 
\author[d,e]{Zhaohuan Zhu}
\author[f,g,h]{Daniel Huber}
\author[i]{Zheng Zheng}
\author[j]{P. De Cat}
\author[k]{J.N. Fu}
\author[a,b]{Hui-Gen Liu}
\author[l]{Ali Luo}
\author[l]{Yue Wu}
\author[l]{Haotong Zhang}
\author[a,b]{Hui Zhang}
\author[a,b]{Ji-Lin Zhou}
\author[m]{Zihuang Cao}
\author[n]{Yonghui Hou}
\author[n]{Yuefei Wang}
\author[n]{Yong Zhang}

\affil[a]{School of Astronomy and Space Science, Nanjing University, Nanjing 210093, China}
\affil[b]{Key Laboratory of Modern Astronomy and Astrophysics in Ministry of
Education, Nanjing University, Nanjing 210093, China}
\affil[c]{Kavli Institute for Astronomy and Astrophysics, Peking University, Yi He Yuan Road 5, Hai Dian District, Beijing 100871, China}
\affil[d]{Department of Astrophysical Sciences, Princeton University, Princeton, NJ 08544}
\affil[e]{Department of Physics and Astronomy, University of Nevada, Las Vegas, NV 89154}
\affil[f]{Sydney Institute for Astronomy (SIfA), School of Physics, University of Sydney, NSW 2006, Australia}
\affil[g]{SETI Institute, 189 Bernardo Avenue, Mountain View, CA 94043, USA}
\affil[h]{Stellar Astrophysics Centre, Department of Physics and Astronomy, Aarhus University, Ny Munkegade 120, DK-8000 Aarhus C, Denmark}
\affil[i]{Department of Physics and Astronomy, University of Utah, Salt Lake City, UT 84112, USA}
\affil[j]{Royal observatory of Belgium, Ringlaan 3, B-1180 Brussel, Belgium}
\affil[k]{Department of Astronomy, Beijing Normal University, 19 Avenue Xinjiekouwai, Beijing 100875, China}
\affil[l]{National Astronomical Observatories, Chinese Academy of Sciences, Beijing 100012, China}
\affil[m]{Key Laboratory of Optical Astronomy, National Astronomical Observatories, Chinese Academy of Sciences, Beijing 100012, China}
\affil[n]{Nanjing Institute of Astronomical Optics \& Technology, National Astronomical Observatories, Chinese Academy of Sciences, Nanjing 210042}

\leadauthor{Xie} 

\significancestatement{The Kepler satellite has made revolutionary discoveries of thousands of planets down to Earth size. However, the orbital shapes (parameterized by eccentricities) of most Kepler planets remain unknown. We derive the eccentricity distributions of an unprecedented large and homogeneous sample of 698 Kepler planets. We discover a dichotomy in eccentricities: the systems with single transiting planets, comprised of half of the sample, have a large mean eccentricity ($\sim$ 0.3) while the multiples are on nearly circular orbits. The average eccentricity and inclination of the Kepler multiples and the Solar System objects fit into an intriguing common pattern. Our results suggest that the circular and coplanar planetary orbits like our Solar System are likely typical in the Galaxy.}

\authorcontributions{J.-W.X. and S.D. designed research; J.-W.X. and S.D. performed research;
J.-W.X. and S.D. contributed new reagents/analytic tools; J.-W.X., S.D., Z. Zhu, D.H., Z. Zheng, A.L., and Y. Wu analyzed data; S.D., P.D.C., J.F., and Haotong Zhang contributed to the planning of Large Sky Area Multi-Object Fiber Spectroscopic Telescope (LAMOST) observations of the Kepler field targets; Z.C., Y.H., Y. Wang, and Y.Z. are LAMOST builders; and J.-W.X., S.D., Z. Zhu, D.H., Z. Zheng, P.D.C., J.F., H.-G.L., A.L., Y. Wu, Haotong Zhang, Hui Zhang, J.-L.Z., Z.C., Y.H., Y. Wang, and Y.Z. wrote the paper.}
\authordeclaration{The authors declare that they have no
competing financial interests}
\equalauthors{\textsuperscript{1}J.-W.X. and S.D. contributed equally to this work.}
\correspondingauthor{\textsuperscript{2}To whom correspondence may be addressed. Email: jwxie@nju.edu.cn or dongsubo@pku.edu.cn.}


\begin{abstract}
The nearly circular (mean eccentricity $\bar{e}\sim 0.06$) and coplanar (mean mutual inclination $\bar{i}\sim 3^{\circ}$) orbits of the Solar System planets motivated Kant and Laplace to put forth the hypothesis that planets are formed in disks, which has developed into the widely accepted theory of planet formation. 
Surprisingly, the first several hundred extrasolar planets (mostly Jovian) discovered using the Radial Velocity (RV) technique are commonly on eccentric orbits ($\bar{e}\sim 0.3$). 
This raises a fundamental question: Are the Solar System and its formation special? 
The {\it Kepler} mission has found thousands of transiting planets dominated by sub-Neptunes, but most of their orbital eccentricities remain unknown.  
By using the precise spectroscopic host star parameters from the LAMOST observations, we measure the eccentricity distributions for a large (698) and homogeneous Kepler planet sample with transit duration statistics.
Nearly half of the planets are in systems with single transiting planets (singles), while the other half are multiple-transiting planets (multiples). 
We find an eccentricity dichotomy: on average, {\it Kepler} singles are on eccentric orbits with $\bar{e}\approx$ 0.3, while the multiples are on nearly circular $(\bar{e} = 0.04^{+0.03}_{-0.04})$ and coplanar $(\bar{i} = {1.4}^{+0.8}_{-1.1}$ degree) orbits similar to the Solar System planets. 
{\xie Our results are consistent with previous studies {\dong of smaller samples} and individual systems.}
We also show that {\it Kepler} multiples and Solar System objects follow a common relation ($\bar{e}\sim$(1-2)$\times\bar{i}$) between mean eccentricities and mutual inclinations.
{\xie 
The prevalence of circular orbits and the common relation may imply that the Solar system is not so atypical in the Galaxy after all.}

\end{abstract}

\dates{This manuscript was compiled on \today}
\doi{\url{www.pnas.org/cgi/doi/10.1073/pnas.XXXXXXXXXX}}

\begin{document}

\verticaladjustment{-2pt}

\maketitle
\thispagestyle{firststyle}
\ifthenelse{\boolean{shortarticle}}{\ifthenelse{\boolean{singlecolumn}}{\abscontentformatted}{\abscontent}}{}

\dropcap{O}ur knowledge of orbital shapes (parameterized with eccentricities) of planetary systems has been drastically advanced in the last two decades largely thanks to 
the RV planet surveys, but there remain some major puzzles.
For example, the majority of RV planets are found on eccentric orbits ($\bar{e}\sim 0.3$)\cite{Wri11} in contrast to the Solar system planets, raising a fundamental question: Is the Solar System an atypical member of the planetary system population in the Galaxy?\cite{Udr07}
Furthermore, the RV method has some key limitations.
For example, several notable biases and degeneracies can introduce considerable systematical uncertainties into the eccentricity distributions derived from the RV technique \cite{ST08, Ang10, Zak11}. 
In addition, the majority of eccentricities measured using the RV method are for giant planets (e.g., Jupiter size), while the eccentricity distributions of smaller planets (e.g., Earth to Neptune size) remain poorly understood. 

Complementary to the RV technique, the {\it Kepler} mission has discovered thousands of planet candidates down to about Earth radius 
using the transit technique \cite{keplercatalog}. About half of the {\it Kepler} planets are in 
systems with multiple transiting planets, and on average they are on 
nearly coplanar orbits similar to the Solar System (see review by \cite{Lis14}).  
For most transiting planets, eccentricities
cannot be directly inferred from the light curves alone. Individual light-curve-based eccentricity measurements have been made for a small number 
of planets, most of which are systems meeting special conditions such as giant planets with high eccentricities\cite{DJ12}, systems with precisely characterized host stars from asteroseismology \cite{VA15} and highly compact and dynamically rich systems exhibiting transit timing variations (TTVs) (e.g. \cite{Lis11a}). Analyzing TTVs for a sample of transit systems also allows to 
constrain eccentricity distributions\cite{HL14}, but this method only applies to a limited number ($\sim 100$) of systems with
special (near-)resonant configurations.

\section*{Method}
A robust general method to derive eccentricity distribution is based on the statistics of transit duration  \cite{For08a} -- the time for transiting planets to  
cross the stellar disks.  Based on Kepler's Third Law, for a planet on a circular orbit that transits across the stellar center, 
the transit duration $T_{0}$ is 
uniquely determined by the orbital period $P$, the planet-to-star radius ratio $r = R_p/R_\star$ and stellar density $\rho_\star$: $T_0 \propto P^{1/3} \rho_{\star}^{-1/3} (1+r)$. For an eccentric and inclined 
orbit, the transit duration $T$ depends on the eccentricity $e$ as 
well as the orientation of the orbit, which is described by the argument of periastron $\omega$ and the impact parameter $b$:
\begin{eqnarray}
T = T_{\rm 0}\times \sqrt{(1-b^2)(1-e^2)}/(1+e{\rm sin}\omega)
\end{eqnarray}
$P$, $T$ and $r$ are observables that can be directly measured from the transit light curve with high precision. 
Using an ensemble of transiting planet systems, the distributions of $\omega$ and $b$ can be modeled and used to infer the eccentricity distribution from the statistics of $T/T_{\rm 0}$. Moreover, for systems with multiple transiting planets, the $b$ distribution also depends on the mutual orbital inclinations {\xie (Throughout this paper, inclination and the symbol $\bar{i}$ always refer to the mutual orbital inclination unless otherwise stated.)} among the planets, and thus both the eccentricity and inclination distributions can be inferred. See {\blue \it{SI Appendix}, section 3,} for a full description of our methodology. 

This method hinges on the well-characterized host properties to 
derive reliable and precise stellar density $\rho_\star$. 
Due to the difficulty of precisely characterizing large samples of stars, the method has hitherto not been applied to a large and homogeneous sample of {\it Kepler} planetary systems. Previous studies such as \cite{Moo11} have used stellar parameters from the {\it Kepler} input catalog (KIC), which are plagued by large systematic uncertainties (see {\blue \it{SI Appendix}, section 1}). A recent study \cite{VA15} uses a sample of 28 multi-transiting systems with precisely measured stellar densities from asteroseismology, and found that the planetary orbits in these systems have low eccentricities. The sample was nevertheless biased towards systems with asteroseismic detections, and did not include single transiting planets.

\section*{Results}

Here, we derive the eccentricity distributions of 698 {\it Kepler} planet candidates using transit duration statistics. The analysis of this large and homogeneous sample \cite{Dec15} is made possible through  spectroscopic observations by the Large Sky Area Multi-Object Fiber Spectroscopic Telescope (LAMOST)\cite{Cui12, Luo15} with reliably derived stellar properties \cite{Don14} from the LAMOST Stellar Parameter (LASP) pipeline (see Sec 4.4 of \cite{Luo15} and also \cite{Wu14}). See {\blue \it{SI Appendix} Section 1} for further discussions on stellar parameters.

\subsection*{Eccentricity Dichotomy}

Our sample consists of 368 systems with single transiting planet {\dong candidates ($N_p=1$)} and 330 {\xie planet candidates} in multiples ($N_p>1$). For the two subsamples, we simulate $T/T_{\rm 0}$ distributions with various $\bar{e}$ and/or $\bar{i}$ assuming Rayleigh distributions, and fit them to the observed $T/T_{\rm 0}$ distributions (see {\blue \it{SI Appendix} Section 4}). The results are shown in Fig.~1 and 2 for the singles and multiples, respectively. 

The eccentricity distribution of singles is clearly non-circular with a mean eccentricity of $\bar{e}=0.32\pm0.02$. In contrast, the orbits of multiples are consistent with being circular and nearly coplanar, with a mean eccentricity $\bar{e}<0.07$ and a mean orbital inclinations 0.006 rad (0.3$^\circ$)$<\bar{i}<$0.038 rad (2.2$^\circ$). We further divide the multiples into subsamples with $N_p=2$, $N_p=3$ and $N_p>3$  transiting planets. We find that the mean eccentricities of these three subsamples are all close to zero and comparable to each other within statistical uncertainties, in contrast to the large mean eccentricity of singles (Fig.~3). Our results indicate an abrupt transition of $\bar{e}$ from singles ($N_p=1$)  to multiples ($N_p>1$) rather than a smooth correlation with $N_p$ as suggested from the study on the RV sample\cite{LT15}. {\xie The low eccentricities of multiples found here are consistent with previous studies of smaller samples\cite{VA15,WL13,HL14} and individual systems, e.g., Kepler-11\cite{Lis11a}, Kepler-36\cite{car12} etc.}

\subsection*{Singles: Two Populations}
When viewed from different orientations, a multiple-planet system can result in multi-transiting and single-transiting systems. Our results provide a direct evidence that not all {\it Kepler} single-transiting system come from the same underlying planet population as the systems with multiple transiting planets. Instead, the {\it Kepler} planet population is likely dichotomic in eccentricities: at least some of the single-transits come from a dynamically hotter population than the underlying population of multi-transiting systems. 

We are therefore motivated to investigate the transit duration ratio of singles with a two-population model (Fig.~4). In the model, we fix the dynamically cold population with an eccentricity distribution corresponding to the best fit of multi-transiting systems, and fit the fraction ($F_{hot}$) and mean eccentricity ($\bar{e}_{hot}$) of the hot population. 
The best-fit two-population model provides a statistically significant improvement over the one-population model in
matching the observations. We find that the hot population makes up a small fraction of the sample ($F_{hot}\sim16\%-36\%$) with an even higher mean eccentricity ($\bar{e}_{hot}>0.47$), 
indicating that the singles are probably dominated by a cold population ($F_{cold}=1-F_{hot}\sim64\%-84\%$).  

\subsection*{Prevalence of Circular Orbits}
As the singles contribute to 
about half of all the {\it Kepler} planets, the fraction of 
planets in the cold population out of the whole sample is even higher ($\sim80\%-90\%$), leading to a conclusion that most {\it Kepler} planets are on near-circular orbits. 
The dominance of near-circular orbits may imply that planets mostly form and evolve in a relatively gentle manner dynamically. 
Violent dynamical scenarios which excite high eccentricities, for example through planet-planet scattering\cite{For08b} and close stellar encounters\cite{Mal07}, therefore must be relatively rare, {\xie unless there are subsequent processes that {\dong efficiently} circularize the orbits.}

\subsection*{A common relation between eccentricity and inclination}
The low eccentricities and inclinations of the {\it Kepler} multiples are naturally expected from simple considerations of terrestrial
planet formation \cite{Tre15}, and they are consistent with the 
expectation from the well-established coplanarity of the {\it Kepler} multiple systems ($\bar{i}\sim$ a few degrees)\cite{TD12, Fab14}, resembling the Solar System planets. 
Further comparing the orbital properties of {\it Kepler} multiples to those of the Solar System objects, we find an intriguing pattern: orbital eccentricities and inclinations are distributed around $\bar{e}\sim$(1-2)$\times\bar{i}$ (see Fig.~5). All the regular moon systems are located in the dynamically cold end (bottom left), while the Asteroid Belt objects and Trans-Neptune objects (TNOs) are located in the dynamically hot end (up-right).
Interestingly, the {\it Kepler} planets and the Solar System planets ($\bar{e}\sim0.06$, $\bar{i}\sim3^\circ$) are located in the intermediate region and close to each other. Note that the {\it Kepler} single-transiting systems are not shown in Fig.~5 because we cannot measure their inclination distributions. If they were following the same pattern, they would be located in the dynamical hot region (up right corner of Fig.~5) given their large mean eccentricity. If true, we would expect that the obliquity of {\it Kepler} single transiting systems should be systematically larger than those of the {\it Kepler} multiples\cite{MW14}.

\section*{Discussions}
In contrast to the low eccentricity of multiple-transit systems, 
the mean eccentricity of single-transit systems is much higher and similar to that of planets found by RV surveys, but unlike the RV planets, most of the {\it Kepler} single-transiting systems are sub-Neptune-sized planets ($<4$ Earth radii). We further compare the singles to the multiples in terms of planetary properties (radius and orbital period) and host properties (stellar mass, radius, metallicity and surface density). We found that these parameters are unlikely to play a decisive role in forming the eccentricity dichotomy ({\blue \it{SI Appendix} section 2.1)}. {\xie The dichotomy may have important implications for planet formation and evolution. From the perspective of evolution, current studies have found that the architecture of a planetary system may depend on various conditions during planet formation, e.g., the total mass and distribution of solid\cite{MB15} and the degree of depletion of gas\cite{Daw16} in the planet-forming disk. From the perspective of evolution, long-term planet-planet interactions can sculpt the planet architectures \cite{Joh12, PW15,Mus15} after planet formation. The eccentricity dichotomy found in this work may help to pin down the initial conditions for planet formation and shed light on planet dynamical evolution.}


{\xie One important concern about the eccentricity dichotomy is the likely larger false positive (FP) rate of the singles as compared to that of the multiples. To investigate this issue, we perform two sets of analyses (see {\blue \it{SI Appendix} section 5.4}). In the first set of analysis, we remove singles with large FP probability to reduce the total FP rate of single sample to a degree (a few percents) that is comparable to that of the multiple sample. In the other analysis, we model the effects of the FP on transit duration ratio distribution by injecting FPs  into our simulations and fit the data. Both analyses lead to results ({\blue \it{SI Appendix} Fig.S13, Fig.S14, Fig.S16 and Fig.S17}) that are consistent with those shown in Fig.~4. Based on these analyses, we conclude that FP should not qualitatively change our main conclusion, namely the Kepler singles are composed of dynamically cold and dynamically hot populations. }

The correlation between eccentricities and inclinations shown in Fig.~5 is generally expected from popular planet formation models, which predict that orbital eccentricities are less than twice of orbital inclinations on average\cite{Ida90} (the grey region in Fig.~5).
{\xie In fact, $\bar{e}\sim2\bar{i}$ is consistent with the prediction of energy equipartition among the various degrees of freedom of planetary orbit\cite{Kok05}}.
The prevalence of circular orbits 
among {\it Kepler} planets and the common pattern between {\it Kepler} multiples and Solar System planets may imply that our planetary system is not so atypical in the Galaxy after all.

\acknow{We thankA. Gould, S. Tremaine, Y. Lithwick, D. Fabrycky,
T. Morton, and Re’em Sari for helpful discussions. J.-W.X. and J.-L.Z. acknowledge
support from the Key Development Program of Basic Research of China (973
Program, Grant 2013CB834900) and the National Natural Science Foundation of
China (NSFC) (Grants 11333002). J.-W.X. is also supported by the NSFC Grant
11403012 and a Foundation for the Author of National Excellent Doctoral Dissertation
of People’s Republic of China. S.D. and J.-L.Z. are supported by the Strategic
Priority Research Program, The Emergence of Cosmological Structures of the Chinese
Academy of Sciences (Grant XDB09000000). S.D. also acknowledges the Project
11573003 supported by NSFC. The LAMOST Fellowship is supported by Special
Funding for Advanced Users, budgeted and administrated by Center for Astronomical
Mega-Science, Chinese Academy of Sciences. J.F. acknowledges the support
of the Joint Fund of Astronomy of NSFC and Chinese Academy of Sciences
through Grant U1231202 and the National Basic Research Program of China (973
Program, Grants 2014CB845700 and 2013CB834900). Y. Wu acknowledges the
NSFC under Grant 11403056. H.-G.L. is supported by the NSFC Grant 11503009.
Hui Zhang is supported by the NSFC Grant 11673011. This work uses the data from
the Guoshoujing Telescope (LAMOST), which is a National Major Scientific Project
built by the Chinese Academy of Sciences. Funding for the project has been provided
by the National Development and Reform Commission. LAMOST is operated
and managed by the National Astronomical Observatories, Chinese Academy of
Sciences. Z. Zhu is a Hubble Fellow.}

\showacknow 


\bibliography{bib}

\begin{thebibliography}{10}

\bibitem{Wri11}
{Wright} JT et~al. (2011) {The Exoplanet Orbit Database}.
\newblock {\em \pasp} 123:412--422.

\bibitem{Udr07}
{Udry} S, {Santos} NC (2007) {Statistical Properties of Exoplanets}.
\newblock {\em \araa} 45:397--439.

\bibitem{ST08}
{Shen} Y, {Turner} EL (2008) {On the Eccentricity Distribution of Exoplanets
  from Radial Velocity Surveys}.
\newblock {\em \apj} 685:553--559.

\bibitem{Ang10}
{Anglada-Escud{\'e}} G, {L{\'o}pez-Morales} M, {Chambers} JE (2010) {How
  Eccentric Orbital Solutions Can Hide Planetary Systems in 2:1 Resonant
  Orbits}.
\newblock {\em \apj} 709:168--178.

\bibitem{Zak11}
{Zakamska} NL, {Pan} M, {Ford} EB (2011) {Observational biases in determining
  extrasolar planet eccentricities in single-planet systems}.
\newblock {\em \mnras} 410:1895--1910.

\bibitem{keplercatalog}
{Mullally} F et~al. (2015) {Planetary Candidates Observed by Kepler. VI. Planet
  Sample from Q1--Q16 (47 Months)}.
\newblock {\em \apjs} 217:31.

\bibitem{Lis14}
{Lissauer} JJ, {Dawson} RI, {Tremaine} S (2014) {Advances in exoplanet science
  from Kepler}.
\newblock {\em \nat} 513:336--344.

\bibitem{DJ12}
{Dawson} RI, {Johnson} JA (2012) {The Photoeccentric Effect and Proto-hot
  Jupiters. I. Measuring Photometric Eccentricities of Individual Transiting
  Planets}.
\newblock {\em \apj} 756:122.

\bibitem{VA15}
{Van Eylen} V, {Albrecht} S (2015) {Eccentricity from Transit Photometry: Small
  Planets in Kepler Multi-planet Systems Have Low Eccentricities}.
\newblock {\em \apj} 808:126.

\bibitem{Lis11a}
{Lissauer} JJ et~al. (2011) {A closely packed system of low-mass, low-density
  planets transiting Kepler-11}.
\newblock {\em \nat} 470:53--58.

\bibitem{HL14}
{Hadden} S, {Lithwick} Y (2014) {Densities and Eccentricities of 139 Kepler
  Planets from Transit Time Variations}.
\newblock {\em \apj} 787:80.

\bibitem{For08a}
{Ford} EB, {Quinn} SN, {Veras} D (2008) {Characterizing the Orbital
  Eccentricities of Transiting Extrasolar Planets with Photometric
  Observations}.
\newblock {\em \apj} 678:1407--1418.

\bibitem{Moo11}
{Moorhead} AV et~al. (2011) {The Distribution of Transit Durations for Kepler
  Planet Candidates and Implications for Their Orbital Eccentricities}.
\newblock {\em \apjs} 197:1.

\bibitem{Dec15}
{De Cat} P et~al. (2015) {Lamost Observations in the Kepler Field. I. Database
  of Low-resolution Spectra}.
\newblock {\em \apjs} 220:19.

\bibitem{Cui12}
{Cui} XQ et~al. (2012) {The Large Sky Area Multi-Object Fiber Spectroscopic
  Telescope (LAMOST)}.
\newblock {\em \raa} 12:1197--1242.

\bibitem{Luo15}
{Luo} AL et~al. (2015) {The first data release (DR1) of the LAMOST regular
  survey}.
\newblock {\em \raa} 15:1095.

\bibitem{Don14}
{Dong} S et~al. (2014) {On the Metallicities of Kepler Stars}.
\newblock {\em \apjl} 789:L3.

\bibitem{Wu14}
{Wu} Y, {Du} B, {Luo} A, {Zhao} Y, {Yuan} H (2014) {Automatic stellar spectral
  parameterization pipeline for LAMOST survey} in {\em Statistical Challenges
  in 21st Century Cosmology}, IAU Symposium, eds.{} {Heavens} A, {Starck} JL,
  {Krone-Martins} A.
\newblock Vol.{} 306, pp. 340--342.

\bibitem{LT15}
{Limbach} MA, {Turner} EL (2015) {Exoplanet orbital eccentricity: Multiplicity
  relation and the Solar System}.
\newblock {\em Proceedings of the National Academy of Science} 112:20--24.

\bibitem{WL13}
{Wu} Y, {Lithwick} Y (2013) {Density and Eccentricity of Kepler Planets}.
\newblock {\em \apj} 772:74.

\bibitem{car12}
{Carter} JA et~al. (2012) {Kepler-36: A Pair of Planets with Neighboring Orbits
  and Dissimilar Densities}.
\newblock {\em Science} 337:556.

\bibitem{For08b}
{Ford} EB, {Rasio} FA (2008) {Origins of Eccentric Extrasolar Planets: Testing
  the Planet-Planet Scattering Model}.
\newblock {\em \apj} 686:621--636.

\bibitem{Mal07}
{Malmberg} D et~al. (2007) {Close encounters in young stellar clusters:
  implications for planetary systems in the solar neighbourhood}.
\newblock {\em \mnras} 378:1207--1216.

\bibitem{Tre15}
{Tremaine} S (2015) {The Statistical Mechanics of Planet Orbits}.
\newblock {\em \apj} 807:157.

\bibitem{TD12}
{Tremaine} S, {Dong} S (2012) {The Statistics of Multi-planet Systems}.
\newblock {\em \aj} 143:94.

\bibitem{Fab14}
{Fabrycky} DC et~al. (2014) {Architecture of Kepler's Multi-transiting Systems.
  II. New Investigations with Twice as Many Candidates}.
\newblock {\em \apj} 790:146.

\bibitem{MW14}
{Morton} TD, {Winn} JN (2014) {Obliquities of Kepler Stars: Comparison of
  Single- and Multiple-transit Systems}.
\newblock {\em \apj} 796:47.

\bibitem{MB15}
{Moriarty} J, {Ballard} S (2015) {The Kepler Dichotomy in Planetary Disks:
  Linking Kepler Observables to Simulations of Late-Stage Planet Formation}.
\newblock {\em ArXiv e-prints}.

\bibitem{Daw16}
{Dawson} RI, {Lee} EJ, {Chiang} E (2016) {Correlations between Compositions and
  Orbits Established by the Giant Impact Era of Planet Formation}.
\newblock {\em \apj} 822:54.

\bibitem{Joh12}
{Johansen} A, {Davies} MB, {Church} RP, {Holmelin} V (2012) {Can Planetary
  Instability Explain the Kepler Dichotomy?}
\newblock {\em \apj} 758:39.

\bibitem{PW15}
{Pu} B, {Wu} Y (2015) {Spacing of Kepler Planets: Sculpting by Dynamical
  Instability}.
\newblock {\em \apj} 807:44.

\bibitem{Mus15}
{Mustill} AJ, {Davies} MB, {Johansen} A (2015) {The Destruction of Inner
  Planetary Systems during High-eccentricity Migration of Gas Giants}.
\newblock {\em \apj} 808:14.

\bibitem{Ida90}
{Ida} S (1990) {Stirring and dynamical friction rates of planetesimals in the
  solar gravitational field}.
\newblock {\em \icarus} 88:129--145.

\bibitem{Kok05}
{Kokubo} E (2005) {Dynamics of planetesimals: the role of two-body relaxation}
  in {\em IAU Colloq. 197: Dynamics of Populations of Planetary Systems},
  eds.{} {Kne{\v z}evi{\'c}} Z, {Milani} A.
\newblock pp. 41--46.

\bibitem{Zha12}
{Zhao} G, {Zhao} YH, {Chu} YQ, {Jing} YP, {Deng} LC (2012) {LAMOST spectral
  survey: An overview}.
\newblock {\em \raa} 12:723--734.

\bibitem{Liu15}
{Liu} XW, {Zhao} G, {Hou} JL (2015) {Preface: The LAMOST Galactic surveys and
  early results}.
\newblock {\em \raa} 15:1089.

\bibitem{Guo15}
{Guo} H et~al. (2015) {Redshift-space clustering of SDSS galaxies - luminosity
  dependence, halo occupation distribution, and velocity bias}.
\newblock {\em \mnras} 453:4368--4383.

\bibitem{Buc12}
{Buchhave} LA et~al. (2012) {An abundance of small exoplanets around stars with
  a wide range of metallicities}.
\newblock {\em \nat} 486:375--377.

\bibitem{Cha14}
{Chaplin} WJ et~al. (2014) {Asteroseismic Fundamental Properties of Solar-type
  Stars Observed by the NASA Kepler Mission}.
\newblock {\em \apjs} 210:1.

\bibitem{Hub13}
{Huber} D et~al. (2013) {Fundamental Properties of Kepler Planet-candidate Host
  Stars using Asteroseismology}.
\newblock {\em \apj} 767:127.

\bibitem{Ren15}
{Ren} JJ et~al. (2015) {On the LSP3 estimates of surface gravity for
  LAMOST-Kepler stars with asteroseismic measurements}.
\newblock {\em ArXiv e-prints}.

\bibitem{Buc14}
{Buchhave} LA et~al. (2014) {Three regimes of extrasolar planet radius inferred
  from host star metallicities}.
\newblock {\em \nat} 509:593--595.

\bibitem{Bro11}
{Brown} TM, {Latham} DW, {Everett} ME, {Esquerdo} GA (2011) {Kepler Input
  Catalog: Photometric Calibration and Stellar Classification}.
\newblock {\em \aj} 142:112.

\bibitem{Dot08}
{Dotter} A et~al. (2008) {The Dartmouth Stellar Evolution Database}.
\newblock {\em \apjs} 178:89--101.

\bibitem{Ser13}
{Serenelli} AM, {Bergemann} M, {Ruchti} G, {Casagrande} L (2013) {Bayesian
  analysis of ages, masses and distances to cool stars with non-LTE
  spectroscopic parameters}.
\newblock {\em \mnras} 429:3645--3657.

\bibitem{SK15}
{Sliski} DH, {Kipping} DM (2014) {A High False Positive Rate for Kepler
  Planetary Candidates of Giant Stars using Asterodensity Profiling}.
\newblock {\em \apj} 788:148.

\bibitem{Sha15}
{Shabram} M, {Demory} BO, {Cisewski} J, {Ford} EB, {Rogers} L (2015) {The
  Eccentricity Distribution of Short-Period Planet Candidates Detected by
  Kepler in Occultation}.
\newblock {\em ArXiv e-prints}.

\bibitem{kane12}
{Kane} SR, {Ciardi} DR, {Gelino} DM, {von Braun} K (2012) {The exoplanet
  eccentricity distribution from Kepler planet candidates}.
\newblock {\em \mnras} 425:757--762.

\bibitem{Zho07}
{Zhou} JL, {Lin} DNC, {Sun} YS (2007) {Post-oligarchic Evolution of
  Protoplanetary Embryos and the Stability of Planetary Systems}.
\newblock {\em \apj} 666:423--435.

\bibitem{JT08}
{Juri{\'c}} M, {Tremaine} S (2008) {Dynamical Origin of Extrasolar Planet
  Eccentricity Distribution}.
\newblock {\em \apj} 686:603--620.

\bibitem{Win10}
{Winn} JN (2010) {Transits and Occultations}.
\newblock {\em ArXiv e-prints}.

\bibitem{Kip10}
{Kipping} DM (2010) {Investigations of approximate expressions for the transit
  duration}.
\newblock {\em \mnras} 407:301--313.

\bibitem{SM03}
{Seager} S, {Mall{\'e}n-Ornelas} G (2003) {A Unique Solution of Planet and Star
  Parameters from an Extrasolar Planet Transit Light Curve}.
\newblock {\em \apj} 585:1038--1055.

\bibitem{Pri15}
{Price} EM, {Rogers} LA, {Johnson} JA, {Dawson} RI (2015) {How Low Can You Go?
  The Photoeccentric Effect for Planets of Various Sizes}.
\newblock {\em \apj} 799:17.

\bibitem{Swi15}
{Swift} JJ et~al. (2015) {Characterizing the Cool KOIs. VIII. Parameters of the
  Planets Orbiting Kepler Coolest Dwarfs}.
\newblock {\em \apjs} 218:26.

\bibitem{FM12}
{Fang} J, {Margot} JL (2012) {Architecture of Planetary Systems Based on Kepler
  Data: Number of Planets and Coplanarity}.
\newblock {\em \apj} 761:92.

\bibitem{Lis12}
{Lissauer} JJ et~al. (2012) {Almost All of Kepler's Multiple-planet Candidates
  Are Planets}.
\newblock {\em \apj} 750:112.

\bibitem{MJ11}
{Morton} TD, {Johnson} JA (2011) {On the Low False Positive Probabilities of
  Kepler Planet Candidates}.
\newblock {\em \apj} 738:170.

\bibitem{Fre13}
{Fressin} F et~al. (2013) {The False Positive Rate of Kepler and the Occurrence
  of Planets}.
\newblock {\em \apj} 766:81.

\bibitem{Des15}
{D{\'e}sert} JM et~al. (2015) {Low False Positive Rate of Kepler Candidates
  Estimated From A Combination Of Spitzer And Follow-Up Observations}.
\newblock {\em \apj} 804:59.

\bibitem{Mor16}
{Morton} TD et~al. (2016) {False Positive Probabilities for all Kepler Objects
  of Interest: 1284 Newly Validated Planets and 428 Likely False Positives}.
\newblock {\em \apj} 822:86.

\bibitem{San12}
{Santerne} A et~al. (2012) {SOPHIE velocimetry of Kepler transit candidates.
  VII. A false-positive rate of 35\% for Kepler close-in giant candidates}.
\newblock {\em \aa} 545:A76.

\bibitem{San15}
{Santerne} A et~al. (2015) {SOPHIE velocimetry of Kepler transit candidates
  XVII. The physical properties of giant exoplanets within 400 days of period}.
\newblock {\em ArXiv e-prints}.

\bibitem{Col12}
{Col{\'o}n} KD, {Ford} EB, {Morehead} RC (2012) {Constraining the false
  positive rate for Kepler planet candidates with multicolour photometry from
  the GTC}.
\newblock {\em \mnras} 426:342--353.

\bibitem{Lis11}
{Lissauer} JJ et~al. (2011) {Architecture and Dynamics of Kepler's Candidate
  Multiple Transiting Planet Systems}.
\newblock {\em \apjs} 197:8.

\end{thebibliography}

\begin{figure*}
\centering
\includegraphics[width=\linewidth]{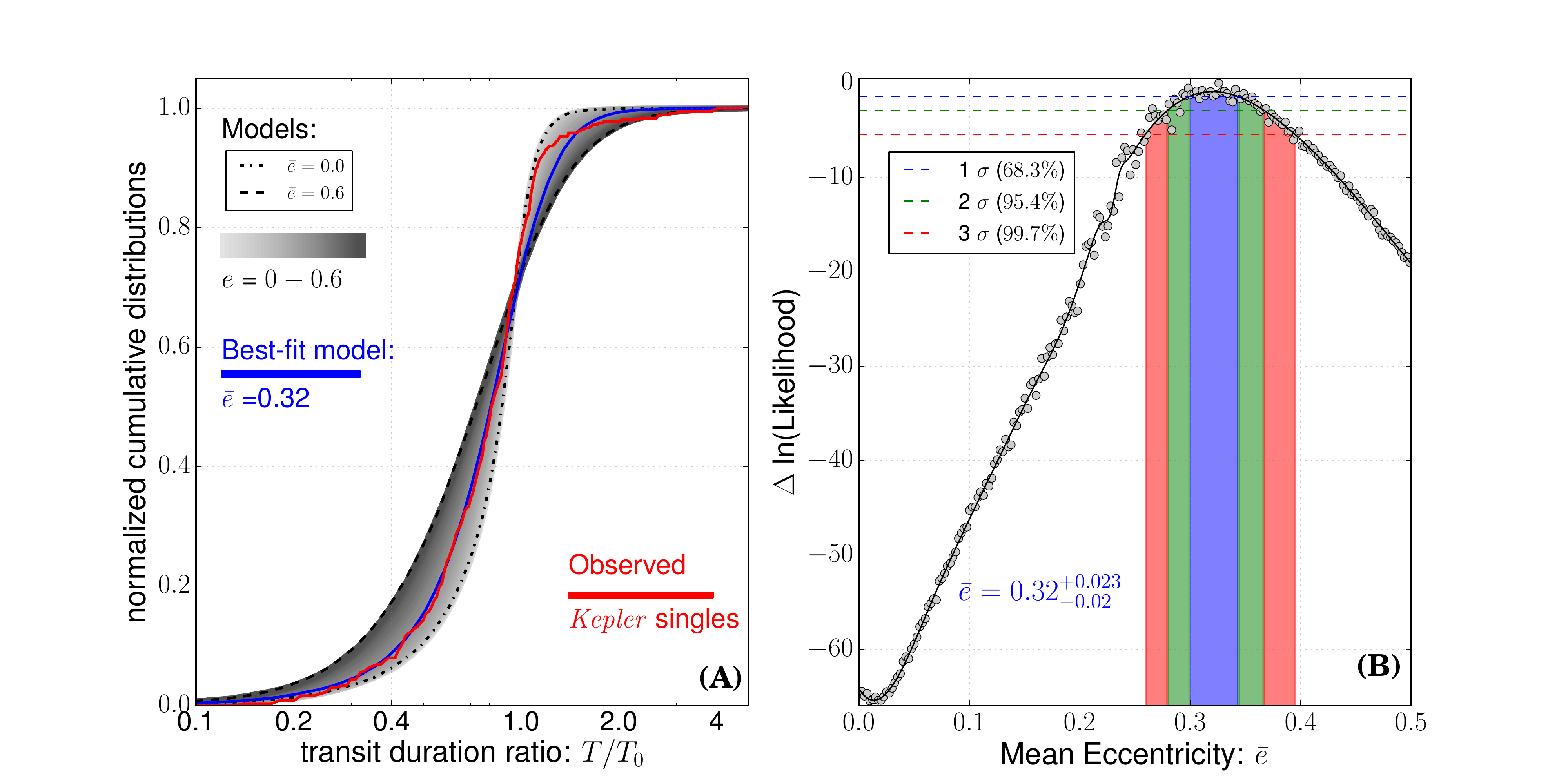}
\caption{{\bf Transit duration ratio statistics of {\it Kepler} single
transiting planets.} {\bf (\textsc{A})} The cumulative distribution of the
observed transit duration $T$ normalized by the expected values for
circular and edge-on orbit $T_0$ for the {\it Kepler} singles (red).
The circular-orbit model (dotted-dashed) is clearly ruled out. The
models shown (grey scale) assume Rayleigh eccentricity distributions
with mean eccentricities $\bar{e}$ varying between 0 and 0.6 (dashed).
The best-fit model with $\bar{e} = 0.32$ is shown in blue. The observed
distribution has relatively large deviations from the best-fit model
in the range of $T/T_0 \sim 1.0 - 2.0$,
which may indicate either the breakdown of the assumed
model Rayleigh distribution or the need for more than one
underlying populations (see Fig.~4 and related discussion in the main text). {\bf (\textsc{B})}
Relative likelihood in logarithm as a function of $\bar{e}$. The blue,
green and red hatched regions indicate the 68.3\% ($1 \sigma$) 95.4\%
($2 \sigma$) and 99.7\% ($3 \sigma$) confidence levels. }
\label{fig_ecc_single}
\end{figure*}

\begin{figure*}
\centering
\includegraphics[width=\linewidth]{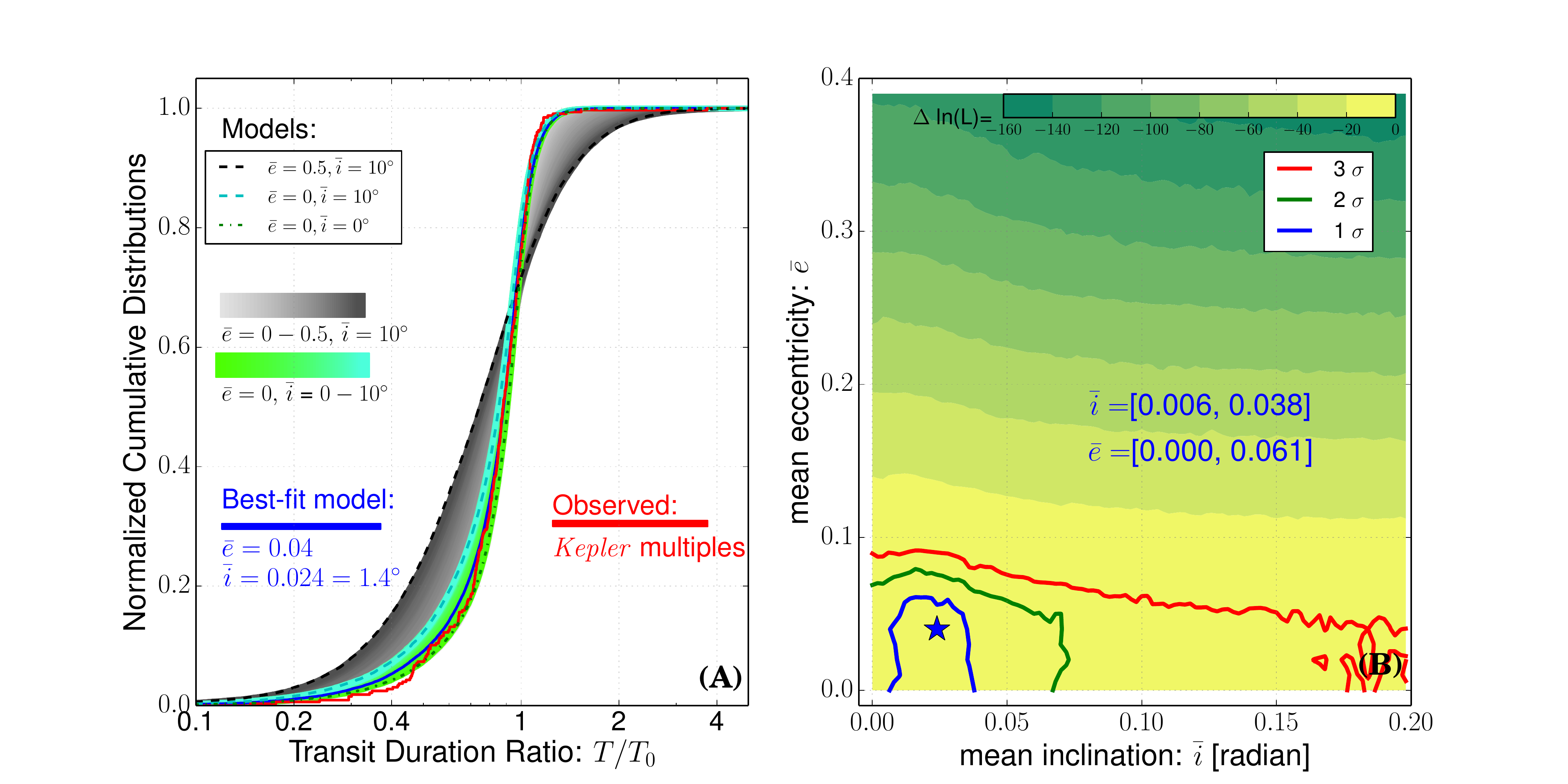}
\caption{{\bf \bf Transit duration ratio statistics of {\it Kepler} multiple
transiting planets.} {\bf (\textsc{A})} Cumulative distribution of observed
transit duration ratios $T/T_0$ for {\it Kepler} multiples is shown in
red. It is fitted
with models assuming
Rayleigh distributions in eccentricities and mutual inclinations. A
range of circular-orbit models with
$\bar{e}=0$ and $\bar{i}$ between 0 and $10^\circ$
are shown in green. The
orbits of {\it Kepler} multiples are consistent with being circular and
nearly coplanar.  The best-fit model (blue) has $\bar{e} = 0.04$ and
$\bar{i} = 0.024 (1.4^\circ)$. For comparison, a range of models
with mean eccentricities between 0 and 0.5
and inclinations between 0 and 10 degrees are shown in grey scale.
{\bf (\textsc{B})} Contours of relative likelihood in logarithm in the
$\bar{i}-\bar{e}$ plane. The blue, green and red contours
indicate the 68.3\% ($1 \sigma$) 95.4\% ($2 \sigma$) and 99.7\% ($3
\sigma$) confidence levels of $\bar{e}$ and $\bar{i}$. The blue star
marks the best-fit values. }
\label{fig_ecc_multiple}
\end{figure*}

\begin{figure*}
\centering
\includegraphics*[width=\textwidth]{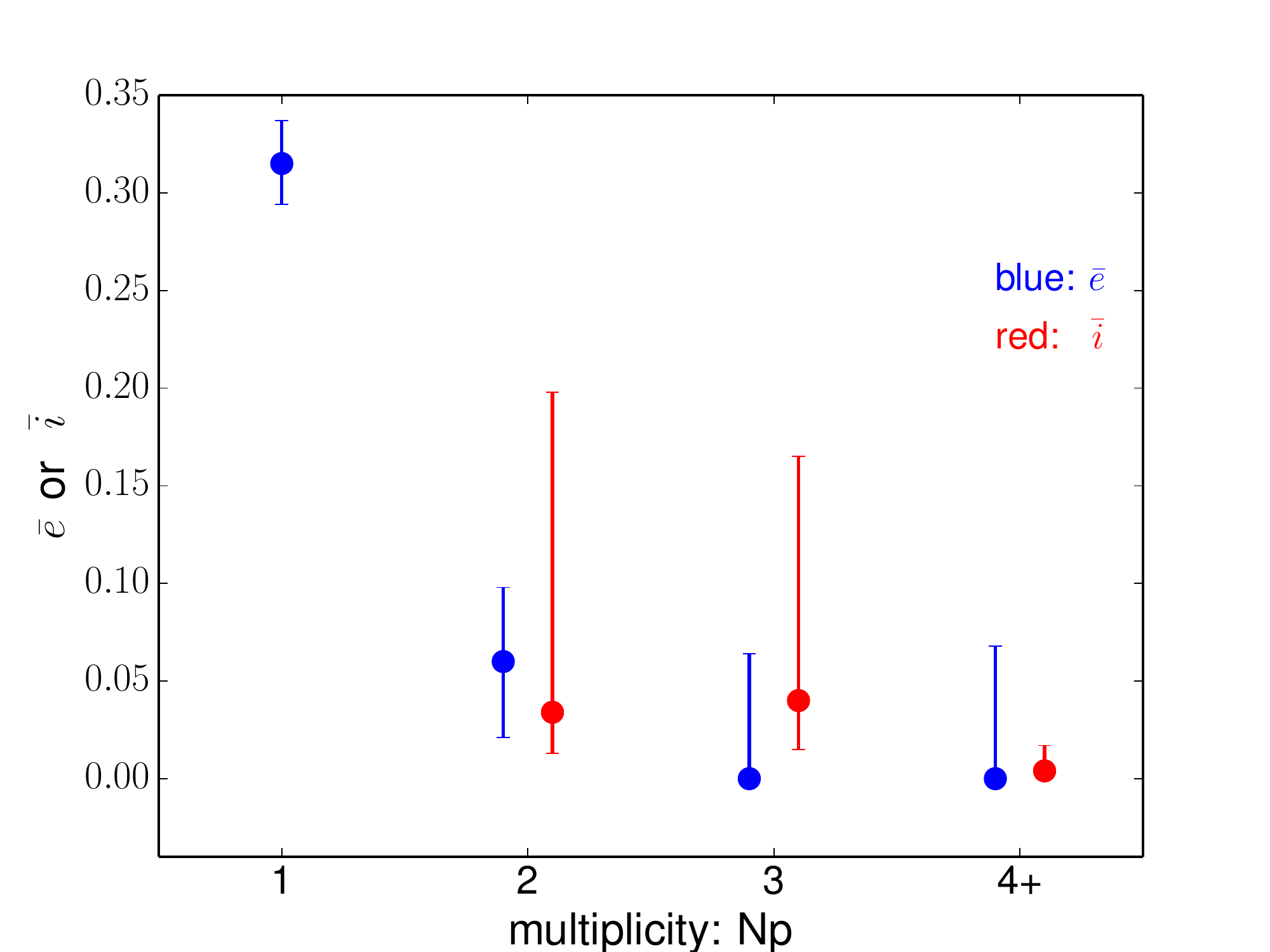}
\caption{{\bf Mean eccentricity $\bar{e}$ (blue markers) and inclination $\bar{i}$ (red markers) as a function of transiting multiplicity $N_p$.} The filled circles show the best fit and the error bars indicate the 68\% confidence interval. As can be seen, it reveals an abrupt transition of $\bar{e}$ rather than a smooth correlation with $N_p$ (see more discussion in {\blue \it{SI Appendix} section 5.1}).} 
\label{fig_ecc_inc_np}
\end{figure*}

\begin{figure*}
\centering
\includegraphics[width=\linewidth]{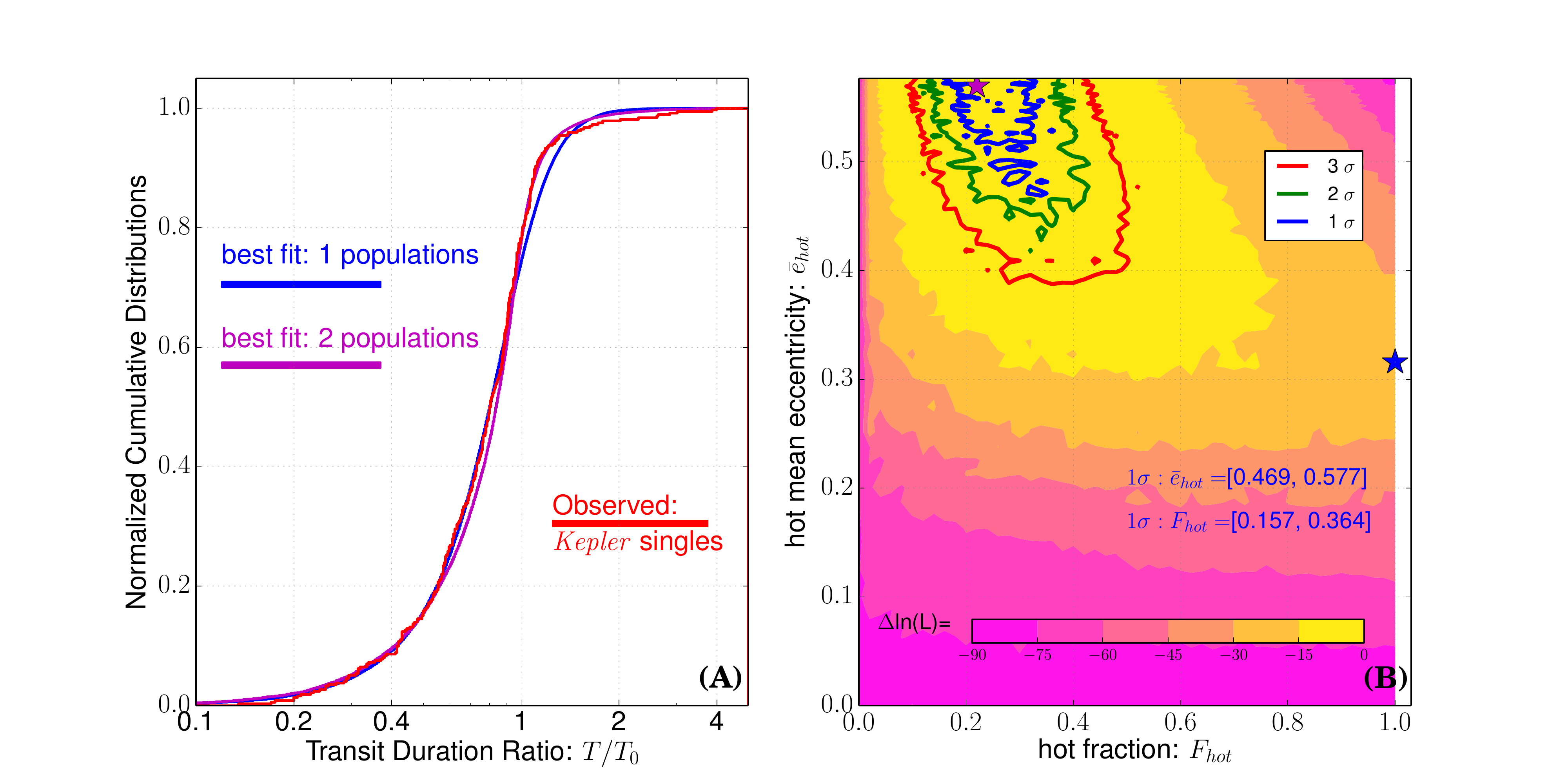}
\caption{{\bf Modeling transit duration ratio distribution ($T/T_{0}$) of single transiting systems with a two-population model.} 
{\bf (\textsc{A})} Similar to the left panel of Fig.~1 except that we model the observed distribution with a two-population model. We fix the dynamical cold population with an eccentricity distribution corresponding to the best fit of multiples as shown in Fig.~2 and set another dynamically hot population with Rayleigh eccentricity distribution with mean $\bar{e}_{hot}$. The fraction ($F_{hot}$) and mean eccentricity ($\bar{e}_{hot}$) of the hot population are two free parameters. {\dong Note that the Rayleigh distribution approaches 
the thermal distribution at $\bar{e} \gtrsim 0.6$.}
{\bf (\textsc{B})} Contours of relative likelihood in logarithm in the
$F_{hot}-\bar{e}_{hot}$ plane. The blue, green and red hatched regions
indicate the 68.3\% ($1 \sigma$) 95.4\% ($2 \sigma$) and 99.7\% ($3
\sigma$) confidence levels. 
The magenta line (left panel) and magenta star symbol (right panel) indicate the best fit of the two-population model, while the blue ones are for the one-population model. The differences in BIC (Bayesian Information Criterion) between the best fits of the one-population and two-population models is $\Delta BIC=-2\Delta {\rm ln({\bf L})}-{\rm ln}(N_{obs})$ = 40.2, which indicates a significant improvement in fitting. }
\label{fig_bimodal}
\end{figure*}

\begin{figure*}
\centering
\includegraphics[width=\linewidth]{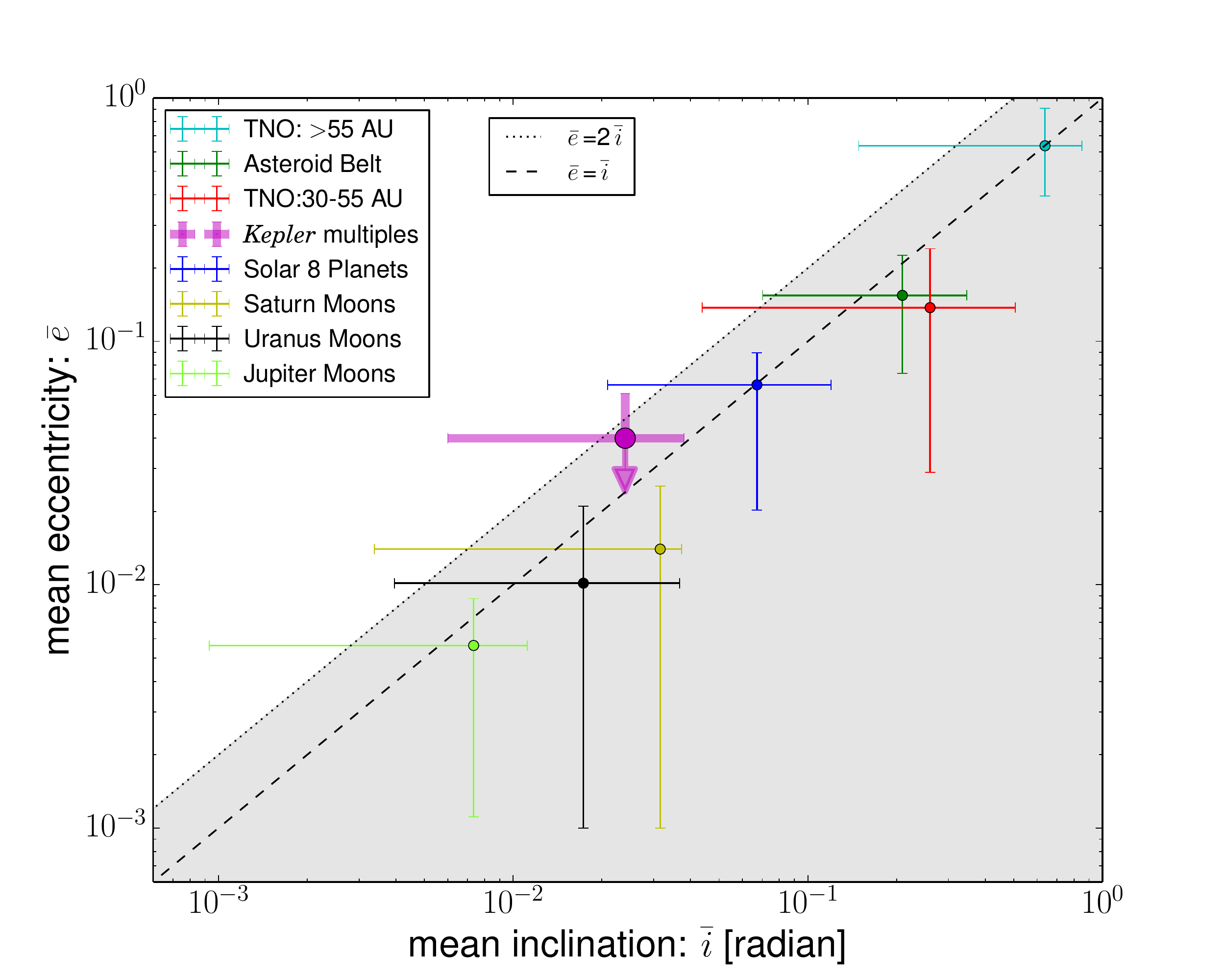}
\caption{{\bf The mean eccentricity and inclination of
{\it Kepler} multiples fit into the pattern of the Solar System objects.}
The thin filled circle and error bars show the mean values and 68\%
confidence intervals of orbital eccentricity and inclination
distributions for  Solar System objects, including 8 planets (blue),
regular moons of Jupiter, Saturn and Uranus (light green, yellow and
black), main belt asteroids (green) and Trans-Neptune objects (TNOs --
both the classical Kuiper Belt objects with orbital semi-major axes 30-55 AU shown in
red and the scattered disk objects with semi-major axes $>$55 AU shown in cyan). They
follow an approximately linear relation with
$\bar{e}\sim$(1-2)$\times\bar{i}$ (dashed and dotted lines).
The thick purple filled circle, error bar (and arrow) show the
eccentricity and inclination constraints of {\it Kepler}
multiples: 0.006 rad (0.3$^\circ$)$<\bar{i}<$0.038 rad (2.2$^\circ$)
and $\bar{e}<0.07$ derived in this work. The {\it Kepler} multiples
fall on the linear relation of the Solar System objects, and they are
on similarly circular and coplanar orbits as the Solar System planets. 
}
\label{fig_inc_ecc}
\end{figure*}

\cleardoublepage

\beginSI
\begin{center}
{ \LARGE  Supporting Information (SI) Appendix }\\[0.5cm]
\end{center}

{\bf \large \center Table of Contents\\}
\noindent
 {\bf (1) Stellar Parameters.} (Fig.S1-S4)\\ \\
 {\bf (2) The Sample.}\\
       2.1 Singles vs. Multiples.  (Fig. S5)\\
       2.2 Comparison with the RV Sample. (Fig. S6)\\
       2.3 Comparison with Previous Studies. (Fig. S7)\\ \\ 
 {\bf (3) Simulations of Transit Duration Ratio Distribution.} (Fig. S8)\\
       3.1 Single Transiting Systems. \\
       3.2 Multiple Transiting Systems.\\ \\
 {\bf(4) Transit Duration Ratio: Observations vs. Simulations.} \\
       4.1 Single Transiting Systems.\\
       4.2 Multiple Transiting Systems.\\ 
 {\bf(5) Further Discussions.} \\
       5.1 An Abrupt Transition or A Smooth Correlation?\\
       5.2 Impact Parameter Distribution. (Fig. S9)\\
       5.3 Singles vs. the Outermost of Multiples. (Fig. S10)\\
       5.4 False Positive.{\xie(Fig. S11-S17)} \\
       5.5 Signal-to-Noise Ratio. \\
       5.6 Stellar Property Calibrator.

\section{Stellar Parameters}

LAMOST has been performing large-scale Galactic surveys with  
the spectral resolution $R\sim 1800$ \cite{Zha12, Liu15}, and the LAMOST observations of the  {\it Kepler} field started in 2011 
\cite{Dec15}. We use the 
stellar parameters extracted from the LAMOST Stellar Parameter Pipeline (LASP) (see Sec 4.4 of \cite{Luo15} and also \cite{Wu14}) in the ``AFGK high quality stellar parameter catalog'' 
of LAMOST DR1, DR2 and DR3-alpha data releases. There are 29553 unique {\it Kepler} targets that have LAMOST/LASP stellar parameters. We perform several internal and external examinations on the accuracy of LAMOST/LASP stellar parameters for the dwarfs.

\begin{figure*}
\centering
\includegraphics*[width=0.9\textwidth]{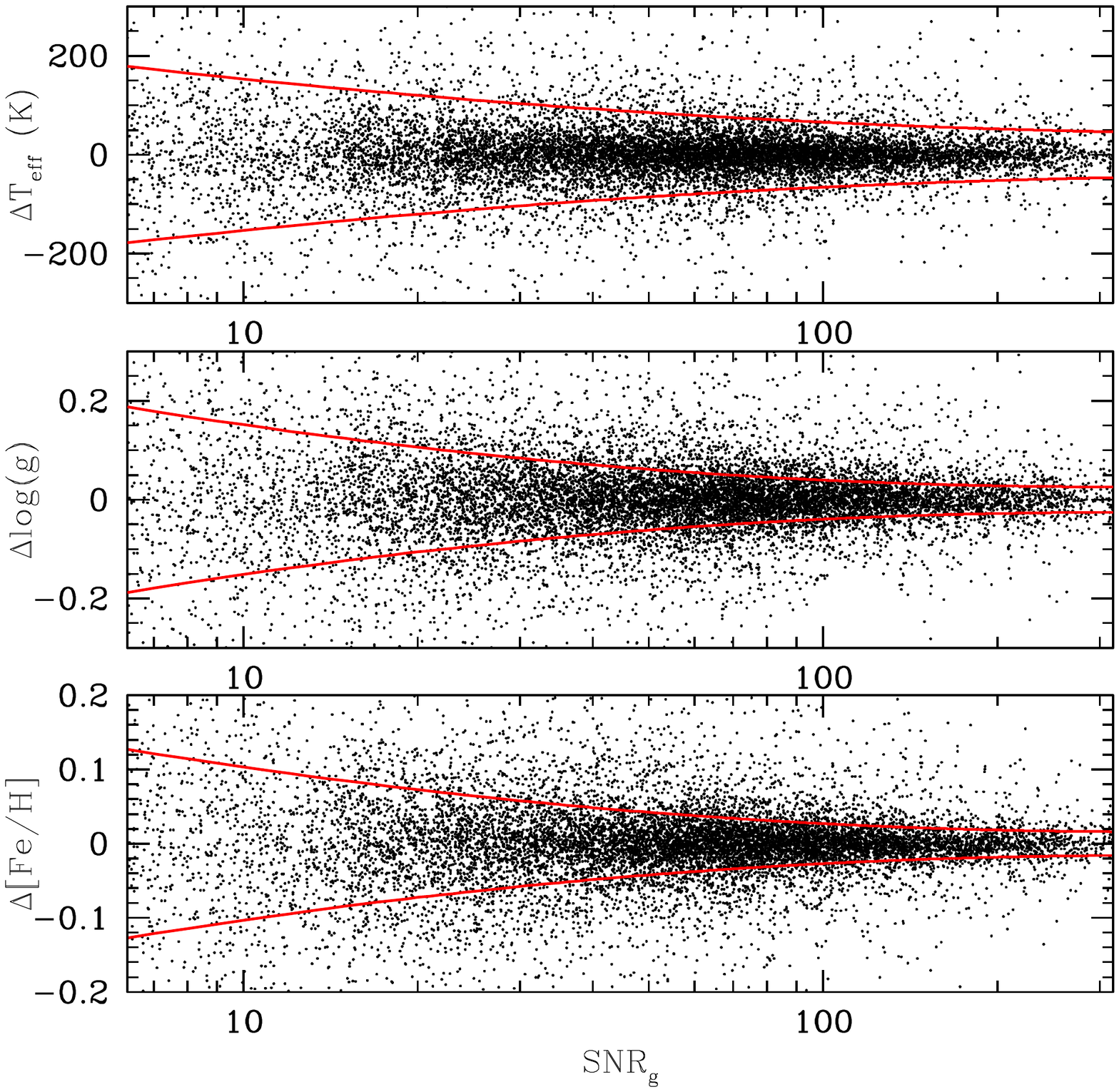}
\caption{{\bf Internal calibration of LAMOST/LASP stellar parameter uncertainties.} The unbiased estimators defined in Supplementary Sec. 1 in $T_{\rm eff}$, $log(g)$, [Fe/H] from the repeated observations as 
a function of SNR$_g$. The $68.3\%$ confidence levels in the {\dong $\log({\rm SNR}_g)$ bins} are 
well described by second-order polynomials {\dong as a function of $\log({\rm SNR}_g)$} plotted in red solid lines. {\dong Note the points are for internal errors and curves describing the internal errors as a function of SNR are later combined with external errors added in quadrature to produce the final error bars (see Fig. \ref{fig_external}).}} 
\label{fig_internal}
\end{figure*}

There are 5924 {\it Kepler} targets that have LAMOST/LASP stellar parameters from more than one epochs of LAMOST observations. 
We assess the internal errors by making comparisons  
of the multi-epoch observations for the same objects. 
We use the unbiased estimator \cite{Guo15}  $\Delta Q_i= \sqrt{n/(n-1)} (Q_i-\bar{Q})$ , with $i=1,2,...,n$, where $i$ denotes each of the individual measurement of $n$ repeated measurements for the stellar parameter $Q$ for each star.
Fig.~\ref{fig_internal} shows $\Delta{T_{\rm eff}}$, $\Delta{log(g)}$ and $\Delta{\rm [Fe/H]}$ using the unbiased estimator (black dots). 
We calculate the $68.3\%$ confidence interval in various bins of $g$-band Signal-to-Noise-Ratio 
per pixel (SNR$_g$) from LAMOST/LASP and find that they 
are well described by the second-order polynomials as a function of 
SNR$_g$ shown in Fig.~\ref{fig_internal}. For measurements with high SNR 
(${\rm SNR}_g > 50$),  the internal errors in $T_{\rm eff}$, $log(g)$ 
and [Fe/H] are less than $35$K, $0.05\,$dex and $0.03\,$dex, respectively.

\begin{figure*}
\centering
\includegraphics*[width=\textwidth]{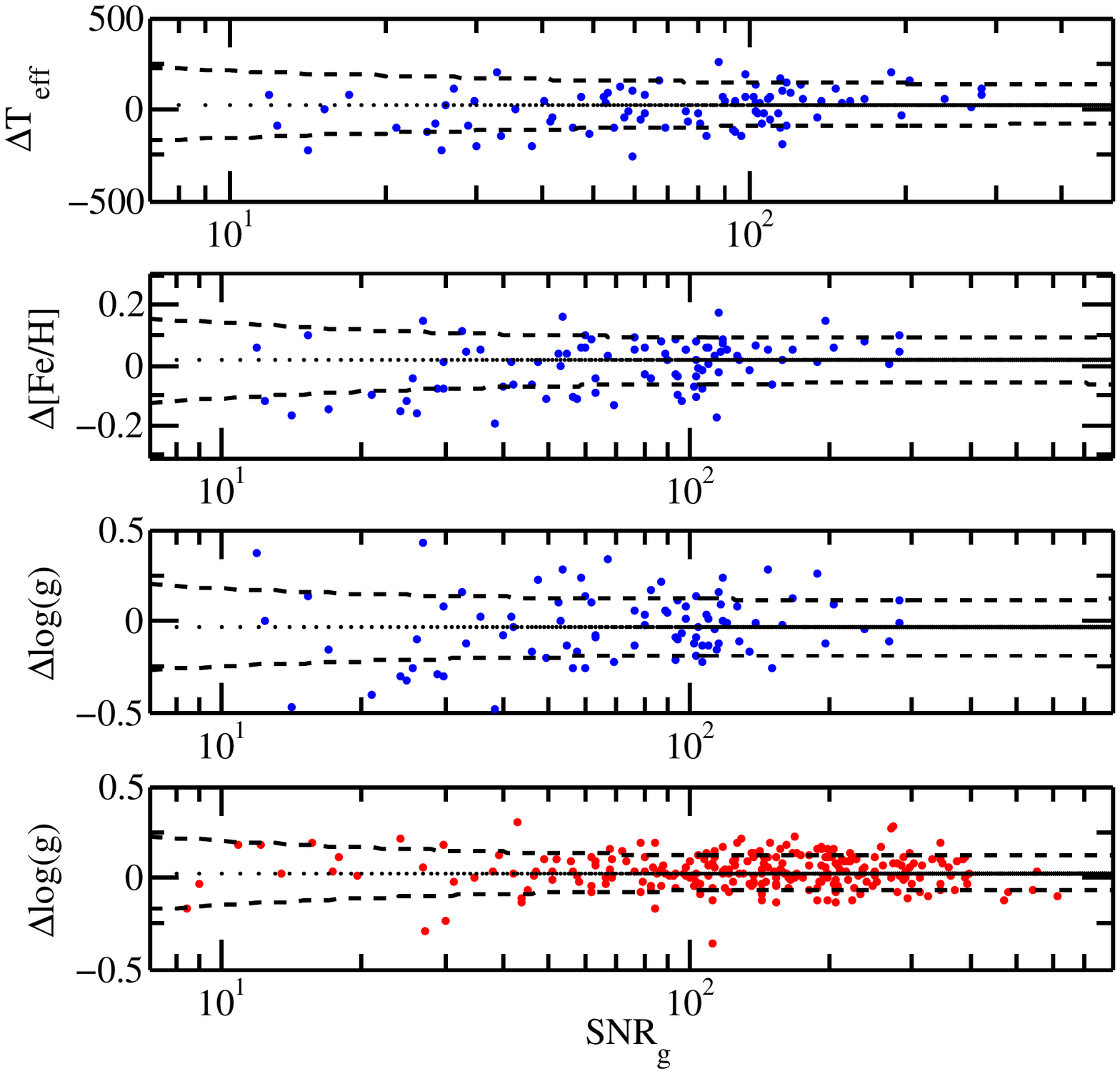}
\caption{{\bf External calibration of LAMOST/LASP stellar parameter uncertainties for dwarfs.} The top 3 panels show comparisons in $T_{\rm eff}$, $log(g)$ and [Fe/H] between LAMOST/LASP and 
SPC method for high-resolution spectra \cite{Buc12} and 
the bottom panel shows the comparison in $log(g)$ with asteroseismology \cite{Cha14}. The dotted line shows the mean 
and the dashed line shows the dispersion taking both 
internal and external uncertainties into account.}
\label{fig_external}
\end{figure*}

Following a previous approach of external examination 
of LAMOST/LASP stellar parameters \cite{Don14}, we compare the 
LAMOST/LASP parameters with those obtained from the 
high-resolution spectroscopy with the
SPC method \cite{Buc12}. There are 87 stars in common between 
the SPC and LAMOST/LASP samples. The results of the comparison  
are shown in the top three panels of Fig.~\ref{fig_external}. 
The mean differences are small: 
${\Delta{T_{\rm eff}}}= 27 K$, ${\Delta{log(g)}}= -0.04 {\,\rm dex}$, ${\Delta{{\rm[Fe/H]}}}=0.015{\,\rm dex}$, 
respectively. For those with ${\rm SNR}_g > 50$, the standard deviations 
in $\Delta{T_{\rm eff}}$, $\Delta{log(g)}$, $\Delta{\rm [Fe/H]}$ are $101 K$, $0.15{\,\rm dex}$, $0.074{\,\rm dex}$, respectively.  Because there are a small
number of common stars with low SNRs, it is difficult to calibrate the 
errors directly from external calibrators for low-SNR measurements.
In order to estimate the error bars for both high- and 
low-SNR measurements, the standard deviations derived from high-SNR measurements are added
in quadrature with the internal error bars in the form of the second-order
polynomials shown in Fig.~\ref{fig_internal}. Since the standard 
deviations for high-SNR measurements from the external calibrations 
are much larger than the internal errors at similar SNRs,  
this approach keeps external error calibrations at high SNRs while 
takes the internal calibrations into account for low-SNR measurements. 

We have also made comparisons in $log(g)$ with the {\it Kepler} asteroseismology 
sample for solar-type stars \cite{Cha14}. There are 260 common stars
between the LAMOST/LASP and the seismology samples. We find that 
the $log(g)$ determinations from LAMOST/LASP are in excellent 
agreement with asteroseismic values, with $\Delta{log(g)}= 0.03 \pm 0.09$ for
${\rm SNR}_{g} > 50$ (see the red 
points in the bottom panel of Fig.~\ref{fig_external}). The dispersion (0.09 dex) is 
smaller than that from the comparison with the spectroscopic sample (0.15 dex). The larger dispersion for the latter likely reflects the 
systematic uncertainties in the SPC method, 
as demonstrated by comparison with the {\it Kepler} seismology sample \cite{Hub13}. We apply the same quadrature corrections taking into 
account for the internal errors and the resulting values as a function 
of SNR are shown in Fig.~\ref{fig_external} (dashed line in the bottom panel).
Note that similar comparisons
have been made before for LAMOST $log(g)$ but 
mostly for giant stars \cite{Ren15} or a mixture of giant and dwarf stars
\cite{Luo15}. For giant stars, LAMOST $log(g)$ appears to have larger uncertainties  
compared to that for the results for the dwarfs studied here.

\begin{figure*}
\centering
\includegraphics*[width=\textwidth]{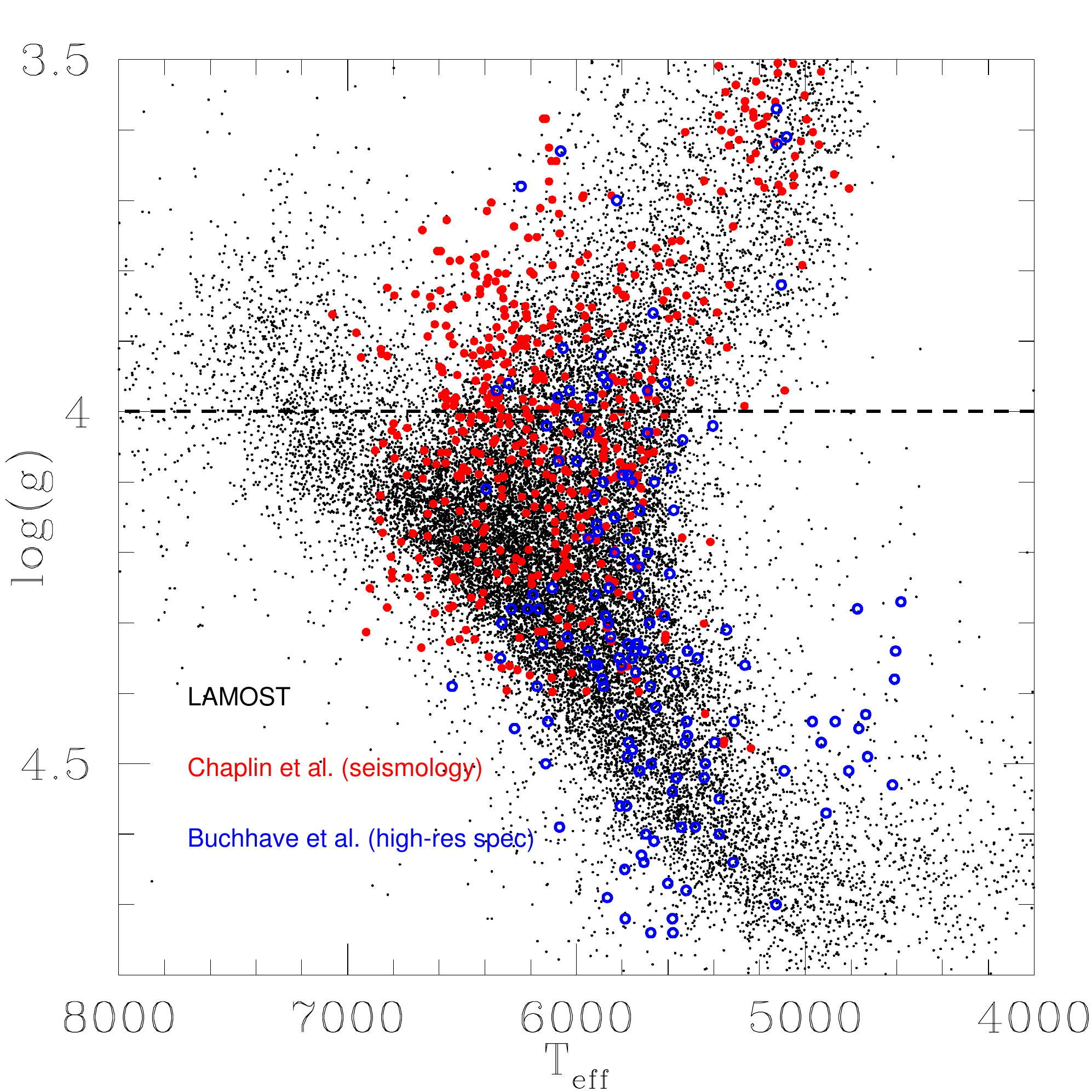}
\caption{{\bf $T_{\rm eff}$ and $log(g)$ distribution for the 
LAMOST/LASP and two external stellar parameter 
calibration samples.} LAMOST stars in the {\it Kepler} field 
are shown in black dots, the asteroseismology sample \cite{Cha14} in 
red solid circles and the high-resolution spectroscopic 
sample \cite{Buc12} in blue open circles. The sample used 
in this work has $log(g)$ larger than $4$, which is shown
in black dashed line.}
\label{fig_teff_logg}
\end{figure*}

Fig.~\ref{fig_teff_logg} shows the $T_{\rm eff}$ and $log(g)$ distributions of the LAMOST (black), high-resolution spectroscopy (blue) and asteroseismology (red) samples discussed above. For the planet hosts studied in our main work, 
we only include dwarfs with $log(g)>4$ ($log(g) = 4$ is shown as dashed line). 
Even though asteroseismology provides higher precision in $log(g)$ than 
high-resolution spectroscopy, the available seismology stars cover poorly for $log(g) > 4.4$ thus possibly limiting the parameter space for its applicability. Given the limitation for both calibrators, we adopt two sets of $log(g)$ with uncertainties determined from high-resolution spectroscopy and seismology respectively, and we derive the eccentricity 
distributions using both sets of $log(g)$ uncertainties separately.
We have also make similar comparisons with the SPC sample published in 2014
\cite{Buc14}, and there are twice as many common stars available as 
compared to the 2012 sample \cite{Buc12} used above.
The mean differences and standard deviations of $\Delta{T_{\rm eff}}, \Delta{log(g)}, \Delta{\rm [Fe/H]}$ are $15 K \pm 111 K, -0.04 \pm 0.15, -0.05 \pm 0.14$ for the 2014 sample. The standard deviations in $\Delta{T_{\rm eff}}$ and $\Delta{log(g)}$ are similar to the 2012 sample while twice
larger in [Fe/H]. This likely due to the new prior in $log(g)$ introduced to 
the 2014 study \cite{Buc14} and the covariance between $log(g)$ and [Fe/H]. In this work, we adopt the uncertainties derived from earlier SPC sample \cite{Buc12}. 

\begin{figure*}
\centering
\includegraphics*[width=\textwidth]{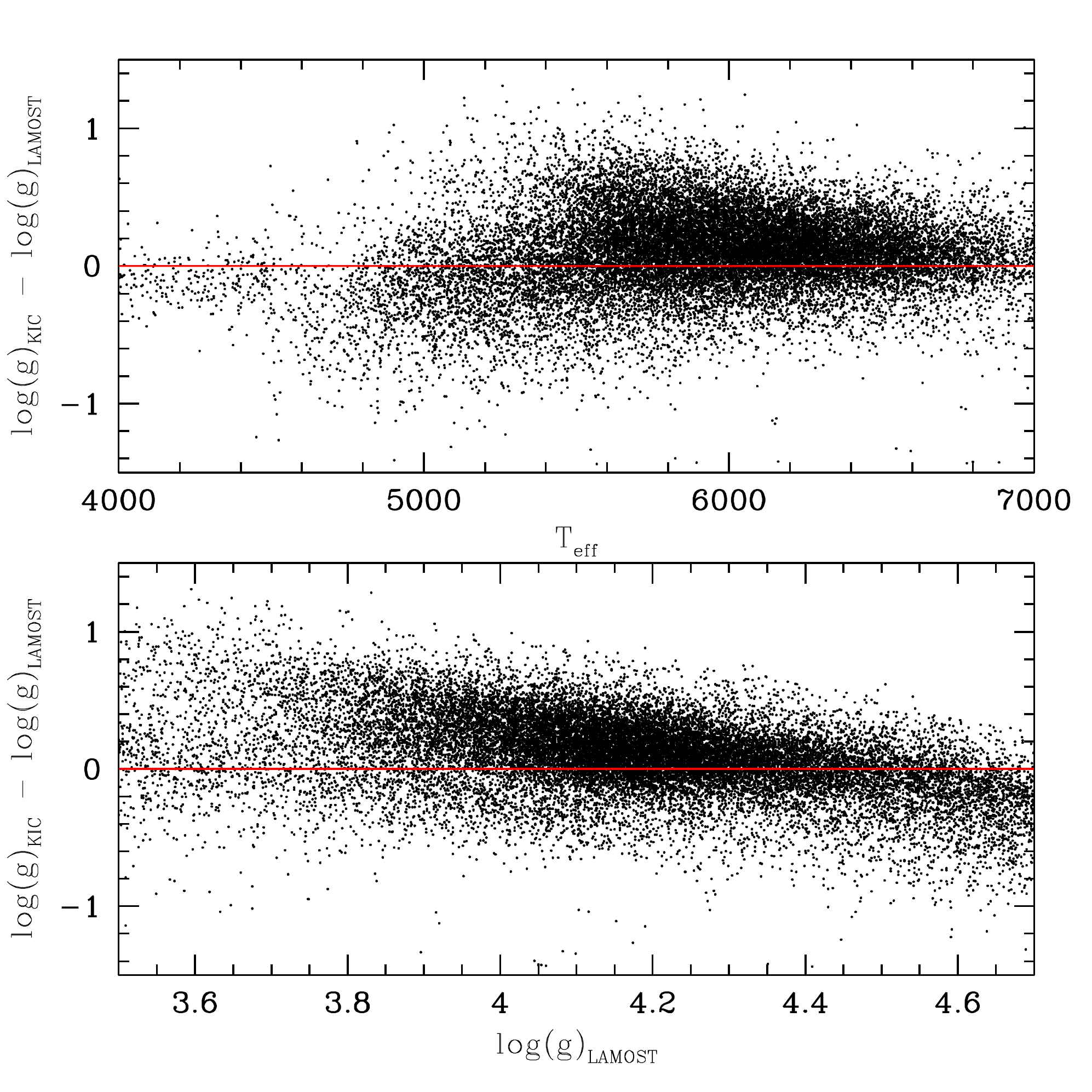}
\caption{{\bf Comparison between KIC and LAMOST $log(g)$ 
reveals large uncertainties and systematics in the KIC determinations}.
The difference between KIC and LAMOST $log(g)$ are shown
as a function of $T_{\rm eff}$ and $log(g)$ determined from 
LAMOST. There are not only large dispersion but also 
large systematic trend, in particular as a function of $log(g)$.}
\label{fig_kic}
\end{figure*}

Fig.~\ref{fig_kic} shows the comparison in $log(g)$ between the {\it Kepler} input
catalog (KIC) \cite{Bro11} and LAMOST/LASP. KIC stellar parameters 
are widely used for studies of {\it Kepler} planets, including previous 
studies of {\it Kepler} planet distributions using transit duration statistics \cite{Moo11}. From the comparison, for stars with $log(g)_{\rm LAMOST}>3.5$, the standard deviation of $\Delta{log(g)} = log(g)_{\rm KIC} - log(g)_{\rm LAMOST}$ is 0.3 dex, translating to 0.45 dex in uncertainties for $\rho_*$.
In addition, there are serious trends of $\Delta{log(g)}$ as a function 
of stellar parameters, in particular $log(g)$. The average $\Delta{log(g)}$
is close to zero for stars with close to solar gravity $log(g) \sim 4.4$, but
for the stars bigger than the Sun, the KIC $log(g)$ values tend to be under-estimated while for the stars smaller than the Sun, the KIC $log(g)$ tend to be over-estimated. The dynamical range of KIC $log(g)$ is smaller than the 
spectroscopic $log(g)$. The large dispersion and severe systematic 
render any statistical studies based on KIC $log(g)$ likely untrustworthy. 

We determine the stellar mass, radius and density with the LAMOST/LASP 
$T_{\rm eff}$, $log(g)$, ${\rm [Fe/H]}$ using isochrone fitting on a dense
grid of isochrones. We use the 2012 version of the interpolated isochrones from ``The Dartmouth Stellar Evolution Database'' \cite{Dot08} with a range of [Fe/H] from -1.5 dex to 0.5 dex
 (grid size of 0.02 dex) and stellar age from 1 to 13 Gyrs (grid size of 0.5 dex). 
We have also applied a separate method \cite{Ser13} with 
the Dartmouth isochrones and found good consistency between the two.

\section{The Sample}
We adopt the transit parameters from the cumulative {\it Kepler} planet candidate catalog 
reported at the NASA Exoplanet Archive (exoplanetarchive.ipac.caltech.edu; retrieved on June 29th, 2015). We crossmatch the {\it Kepler} planet candidates with LAMOST data releases discussed above, and there are 941 planet candidates with host stars characterized by the LAMOST spectroscopy. We further rule out those unusual large candidates (radius $R_p>$ 15 Earth radii) and those with too low transit signal noise ratio (SNR$<7.1$) (more discussion on the cuts on planet parameters in Section {\xie 5.4 and 5.5}) and sub-giant/giant hosts (surface gravity $log(g)<$4), which have relatively large false positive rate\cite{SK15}. We also rule out a handful of  candidates with less than 3 transit, as the number of transit is too few to obtain accurate orbital period. After the cuts, we have a final sample of 698 planet candidates orbiting 501 stars. The stellar and planetary properties can be accessed from the Supplementary dataset. 

\subsection{2.1 Singles vs. Multiples}
\begin{figure*}
\centering
\includegraphics*[width=\textwidth]{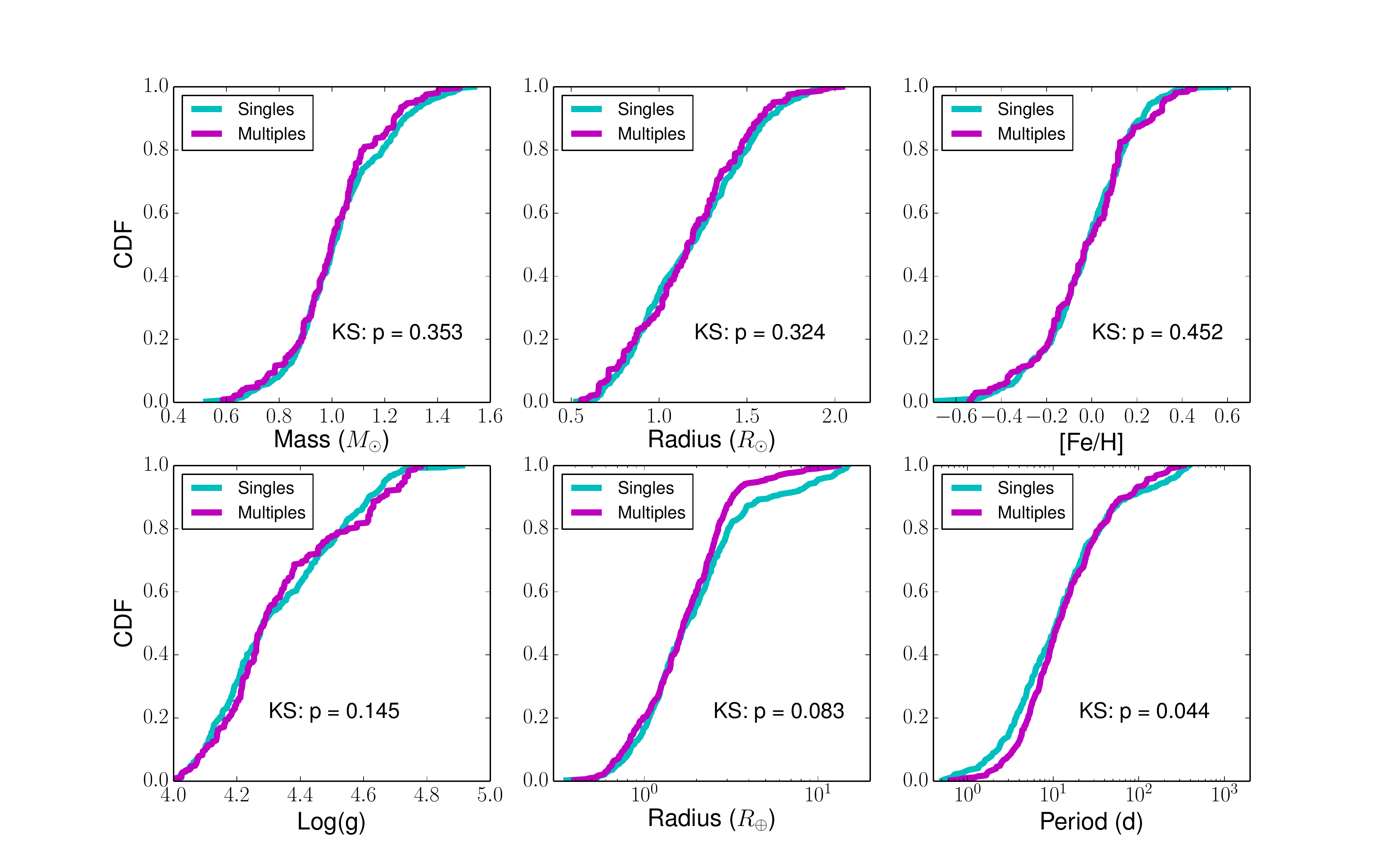}
\caption{{\bf Normalized cumulative distributions of stellar (mass, radius, metallicity and surface gravity) and planetary (radius and orbital period) properties}. The singles in our sample are shown in cyan and multiples are 
shown in magenta. In each panel, the $p$ value of the two-sample KS test for the distributions of singles and multiples is shown at bottom right.  (See more discussions in Supporting Information Sec. 2.1)}
\label{fig_sin_mul}
\end{figure*}

Our sample consists of 368 and 330 planet candidates in single and multiple transiting systems, respectively.  The multiplicity rate is about $47\%$, which is comparable to that of the total {\it Kepler} sample ($42\%$ if applying the same cuts above), showing that LAMOST observations were unbiased with respect to singles or multiples as compared 
to the full {\it Kepler} sample. In Fig.~\ref{fig_sin_mul}, we compare various properties (stellar and planetary) for the two subsets. All stellar parameters are from the LAMOST spectral characterizations. The planet orbital periods are adopted from the {\it Kepler} catalog. The planet radii are calculated via $R_{p}=R_\star*r$ with $R_\star$ adopted derived from LAMOST and the ratio of planet and star radius, $r$, from the {\it Kepler} catalog. There appears to be no significant difference (i.e., KS p value $<$5\%) between the singles and multiple in terms of stellar mass, radius, metallicity, surface gravity, except for planetary radius and orbital period. If we further cut the sample to eliminate those candidates in the regime with relatively high false positive rate  (i.e., $R_p>6R_\oplus$ and orbital period $P<$ 3 day), we find that the differences in various parameters between singles and multiples are even smaller. As we show below (Section 4 and 5), these two populations however differ substantially in their transit duration ratio distributions and thus their orbital eccentricities.

\subsection{2.2 Comparison with the RV Sample}
\begin{figure*}
\centering
\includegraphics*[width=\textwidth]{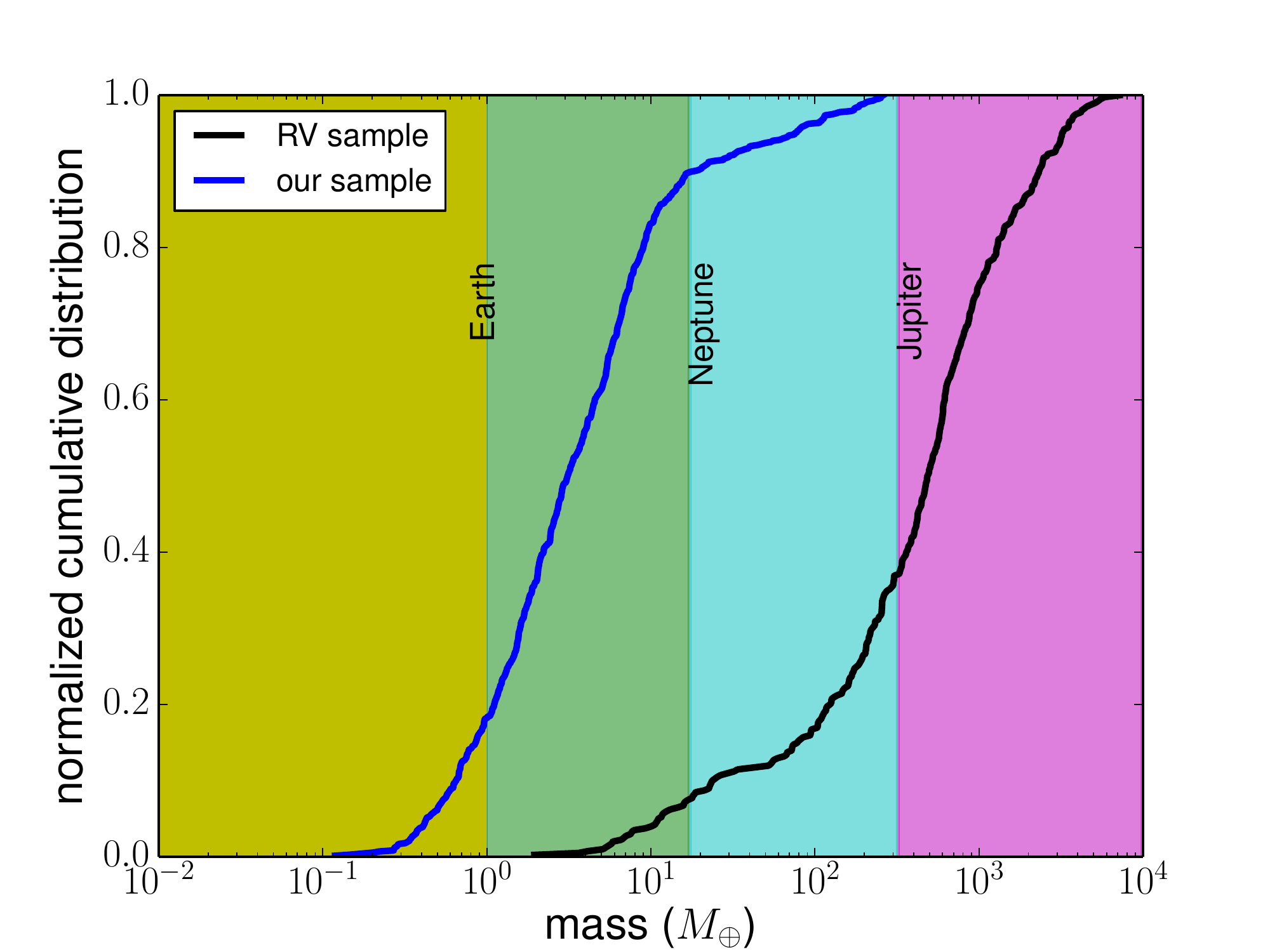}
\caption{{\bf Normalized cumulative distributions of planetary masses of RV planets and the {\it Kepler} sample (blue).} For RV planets (black), we show their minimum masses ($M \sin i$). For the {\it Kepler} planet candidates in our samples (blue), we estimate their masses using the mass-radius relation\cite{Lis11}: $M/M_\oplus=(R/R_\oplus)^{2.06}$.  The RV sample mainly consists of Jupiter-sized giant planets, while our sample is dominated by small planets from Earth to super-Earths or sub-Neptunes.}
\label{fig_kep_rv}
\end{figure*}

From the Exoplanet Orbit Database (exoplanets.org) \cite{Wri11}, we find 439 RV planets. Fig.~\ref{fig_kep_rv} shows the distributions of planetary mass in the RV sample and the estimated mass assuming a simple
mass-radius relation from our sample. We see that the two samples occupy different part of parameter space. The RV sample is primarily composed of giant planets, while our sample contains mainly small planets (Earth to super-Earths and/or sub-Neptunes). 

\subsection{2.3 Comparison to Previous Studies}
Recently, Hadden \& Lithwick (2014)\cite{HL14} have extracted the eccentricity distribution of 139 near-resonance {\it Kepler} planets/candidates from transit timing variation (TTV sample). They find the orbits of these near-resonance planets are nearly circular with mean eccentricity $\bar{e}\sim0.02$. Van Eylen \& Albrecht (2015)\cite{VA15} have derived the orbital eccentricities of 66 {\it Kepler} planets (candidates) in 28 multiple transiting systems whose host stars are characterized by asteroseismology (seismology sample). They also find these multiple systems are generally with small eccentricities.  Shabram et al. 2015\cite{Sha15} have calculated the eccentricities of 50 short period {\it Kepler} planets (candidates) by occultation (i.e., secondary eclipse) (occultation sample). They find that the mean eccentricity is about 0.08 and a two-component model provides a better fit.  

\begin{figure*}
\centering
\includegraphics*[width=\textwidth]{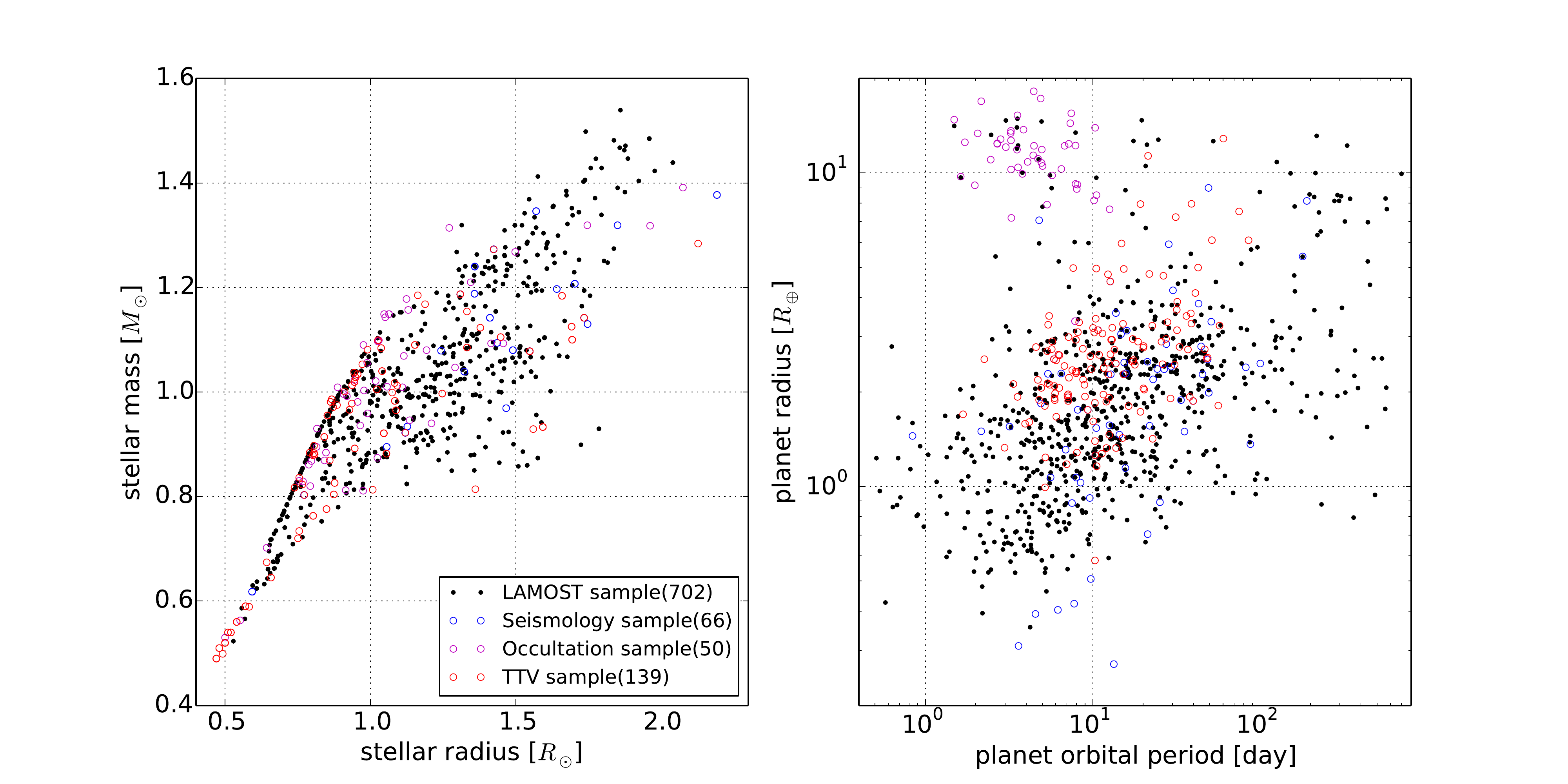}
\caption{{\bf Comparisons in host star (Left Panel) and planet (Right Panel)  properties with 
previous studies.} The LAMOST (this work; black), Seismology (Van Eylen \& Albrecht 2015\cite{VA15}; blue), Occultation (Shabram et al. 2015\cite{Sha15}; magenta) and TTV (Hadden \& Lithwick 2014; red) samples are shown. The numbers in the brackets give the sample size. (See more discussions in Supporting Information Sec. 2.3). }
\label{fig_yoram_simon_megan}
\end{figure*}

In Fig.~\ref{fig_yoram_simon_megan},  we compare our sample (LAMOST sample) to the samples of these previous studies, in terms of stellar and planetary properties. The stars in the seismology sample are biased toward sub-giants. Due to the requirements for having occultation, the occultation sample contains mostly giant planets on 
short orbital period, in contrast with the predominantly sub-Neptune planets of the whole {\it Kepler} planet sample and our sample. We note that  \cite{kane12} also studied the eccentricity distribution of {\it Kepler} giant planets and found that it was consistent with that derived from RV giant planets. Due to the TTV detection limit, the TTV sample are restricted to planets with relatively large radius (larger radius corresponds to higher SNR) and intermediate period (shorter period corresponds to shorter transit duration, longer period corresponds to fewer transit). Furthermore,
TTV studies are restricted to special orbital configurations of 
near-resonance. In contrast, the host stars and planets in the LAMOST sample are broadly distributed 
and represent an unbiased and homogeneous sample from {\it Kepler}.

\section{Simulations of Transit Duration Ratio Distribution}
In this Section, we describe the method and procedure that are used to model the transit duration ratio distribution. Compared to previous studies\cite{Moo11},  there are two major improvements in our modeling.
Moorhead et al. \cite{Moo11} stressed the importance of uncertainties of stellar and transit parameters in modeling transit duration ratio but they did not take
these uncertainties into account when comparing with observation.
We include the uncertainties in our modeling. 
Second, we treat singles and multiples separately. In particular, 
when modeling the transit duration distribution for 
multi-transiting systems, it is important to take
the mutual inclination distribution into account. We stress that these 
two corrections are crucial for transit duration statistics to infer eccentricity from {\it Kepler}, and without taking them into account, 
it can lead to serious errors.

\subsection{3.1 Single Transiting Systems}
For each planet candidate in single-transiting systems, we perform the following steps to generate a simulated transit duration ratio. 

\emph{Step 1:} We first draw an eccentricity ($e$) between 0 and 1 from a Rayleigh distribution\cite{Zho07,JT08}, 
\begin{eqnarray}
dN = \frac{e}{\sigma_e^2}{\rm exp}\left(-\frac{e^2}{2\sigma_e^2}\right)de,
\label{dne}
\end{eqnarray} 
where $\sigma_e$ is the Rayleigh parameter and the mean eccentricity $\bar{e}=\sigma_e\sqrt{\pi/2}$. We repeat this step if $(1-e)a_{\rm R}<1$ because it is unphysical (the planet is inside the star). Here $a_{\rm R}$ is the ratio of orbital semi-major axis ($a$) and the radius of the host star ($R_\star$). Considering the Kepler's third law, we have 
\begin{eqnarray}
a_{\rm R}=4.2\left(\frac{P}{\rm d}\right)^{2/3}\left(\frac{\rho_\star}{\rho_\odot}\right)^{1/3},
\end{eqnarray}
where $P$ and $\rho_\star$ are the transit orbital period and the density of the host star, and they are adopted from the observed values for a given system. Note that for large $\sigma_e$, due to the eccentricity cutoffs ($e<1$ and $(1-e)a_{\rm R}>1$), the mean eccentricity of those drawn from our simulation is smaller than  $\bar{e}=\sigma_e\sqrt{\pi/2}$, and we report the mean eccentricity calculated from the simulation.

\emph{Step 2:} We draw a prior argument of pericenter ($\omega$) and a prior $\cos i_0$ from a uniform distributions and calculate the impact parameter, 
\begin{eqnarray}
b = a_{\rm R}\varrho_c{\rm cos}i_0,
\label{imp}
\end{eqnarray} 
where $i_0$ is the orbital inclination with respect to the plane of the sky, and
\begin{eqnarray}
\varrho_c = (1-e^2)/(1+e{\rm sin}\omega).
\end{eqnarray}
If $b>1+r$, where 
\begin{eqnarray}
r=R_{p}/R_{\star}
\end{eqnarray}
 is the radius ratio of planet/star, it indicates that no transit occurs, then we go back to Step 1. In practice, we set the maximum of $\cos i_0$ as ${\bf min}(1, (1+r)/(a_R\varrho_c))$ to avoid drawing a lot of non-transiting cases. {After the above transit selection (by cutting off $b>1+r$), the distribution of the simulated $\omega$ (especially in the case of large eccentricity) will deviate from the prior uniform distribution because of the well-known geometric effect\cite{Win10}, namely, transit probability is enhanced near the periastron and reduced near the apstron.}

\emph{Step 3:} We calculate the transit duration (total duration, first to fourth contact) following Kipping (2010)\cite{Kip10}, namely
\begin{eqnarray}
T = \frac{P}{\pi}\frac{\varrho_c^2}{\sqrt{1-e^2}} {\rm arcsin}\left(\frac{\sqrt{(1+r)^2-a_{\rm R}^2\varrho_c^2{\rm cos^2}i_0}} {a_{\rm R}\varrho_c{\rm sin}i_0}\right).
\label{tdur}
\end{eqnarray}
For {\it illustrating purpose only (we always use Equation \ref{tdur} in our computation)}, the above equation can be approximately reduced to, 

\begin{eqnarray}
T \sim T_{\rm 0} *F(e,\omega,b),
\end{eqnarray}
where the function
\begin{eqnarray}
F(e,\omega,b)=\frac{\sqrt{(1-b^2)(1-e^2)}}{1+e{\rm sin}\omega}, 
\end{eqnarray}
and $T_0$ is a characteristic time scale, denoting the transit duration if the planet moves on a circular orbit and transits the center of the host star. $T_0$ is calculated from Equation \ref{tdur} by setting $e=0$ and $b=0$, and it can be expressed as 
 \begin{eqnarray}
T_{\rm 0} = 13\, {\rm hr} (P/{\rm yr})^{1/3} (\rho_{\star}/\rho_{\odot})^{-1/3}(1+r).
\end{eqnarray}

The transit duration ratio (TDR) is defined by 
\begin{eqnarray}
{\rm TDR}_{\rm mod}  = T/T_{\rm 0} .
\label{tdrmod}
\end{eqnarray}
Note that here ${\rm TDR}_{\rm mod} $ is the duration ratio predicted from pure theoretical model without any observational uncertainty.

\emph{Step 4:} To simulate a transit duration ratio observation, one needs to consider the observational uncertainties of $T$ and  $T_{\rm 0}$. The simulated duration $T_{\rm sim}$ is obtained by taking the observed duration into account (assuming a Gaussian distribution), 
\begin{eqnarray}
T_{\rm sim} = T*[1+R_N*(\sigma_{T_{obs}}/T_{\rm obs})],
\label{Tsim}
\end{eqnarray}
where $\sigma_{T_{obs}}$ is the observed uncertainty of $T_{\rm obs}$ and $R_N$ is a random variable drawn from a normal distribution centered at 0 and with standard deviation of 1.
Here we assume the simulated transit duration has the same {\it relative} uncertainty as the observation.

\begin{figure*}
\centering
\includegraphics*[width=\textwidth]{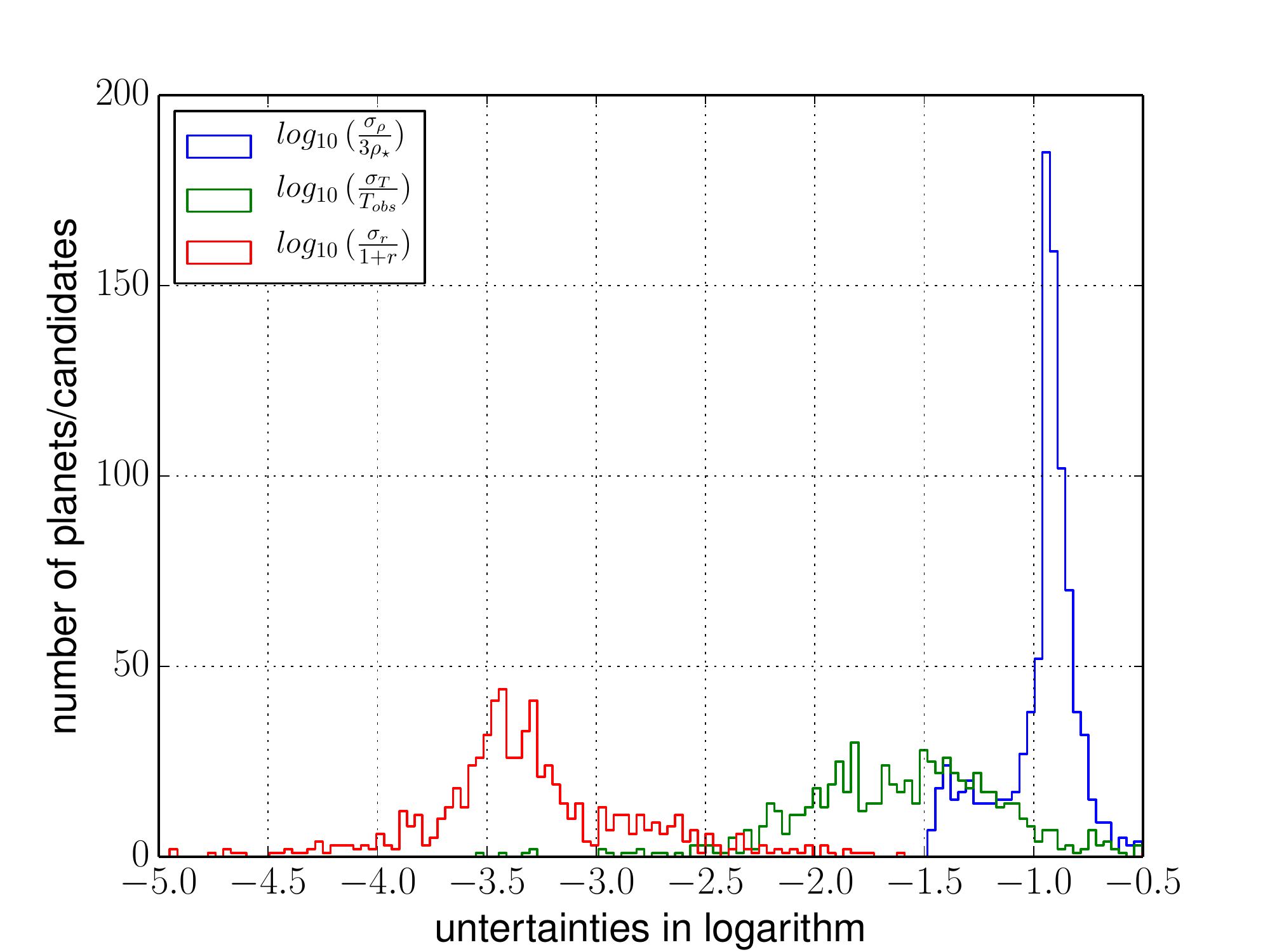}
\caption{{\bf Three sources uncertainty (equation \ref{tdrsig}) from the measurements of transit duration $\frac{\sigma_T}{T_{obs}}$, stellar density $\frac{\sigma_\rho}{3\rho_{\star}}$ and planet/star radius ratio $\frac{\sigma_r}{1+r}$ }.} 
\label{fig_errors}
\end{figure*}

Similarly,
\begin{eqnarray}
T_{0\rm sim} = T_0*[1+R_N*(\sigma_{T_0}/T_0)].
\label{T0sim}
\end{eqnarray}
By propagating errors, the relative error of $T_0$ can be expressed as
\begin{eqnarray}
\sigma_{T_0}/T_0 = \sqrt{\left(\frac{\sigma_\rho}{3\rho_\star}\right)^2+ \left(\frac{\sigma_r}{1+r}\right)^2}
\label{eT0}
\end{eqnarray}
where $\sigma_\rho$ and $\sigma_r$ are the observed uncertainties of $\rho_\star$ and $r$, and $R_N$ is a random number drawn from a normal distribution centered at 0 and with standard deviation of 1. Note the two $R_N$ in equations (\ref{Tsim}) and (\ref{T0sim}) are indepedent.  For each simulated $T_{\rm sim}$, following Fabrycky et al.\cite{Fab14}, we assign a signal noise ratio SNR$_{\rm sim}=$ SNR$_{\rm obs} \sqrt{T_{\rm sim}/T_{\rm obs}}$, where SNR$_{\rm obs}$ and $T_{\rm obs}$ are the observed transit signal noise ratio and duration. We uses a SNR cut to take into account 
the detection efficiency of the {\it Kepler} pipeline. We go back to Step 1 if SNR$_{\rm sim}<7.1$. {About 5\% of simulated transits are eliminated by the SNR cut.} Otherwise, the ratio $T_{\rm sim}/T_{0\rm sim}$ contributes to the simulated transit duration ratio distribution, namely,
\begin{eqnarray}
{\rm TDR}_{\rm sim}  = T_{\rm sim}/T_{\rm0sim}. 
\end{eqnarray}
 The uncertainty of ${\rm TDR}_{\rm sim}$ or ${\rm TDR}_{\rm obs}$ is given by 
\begin{eqnarray}
\sigma{\rm TDR} = \sqrt{\left(\frac{\sigma_{T_{obs}}}{T_{\rm obs}} \right)^2+\left(\frac{\sigma_\rho}{3\rho_\star}\right)^2+ \left(\frac{\sigma_r}{1+r}\right)^2}.\label{tdrsig}
\end{eqnarray}
{Here, the covariances among the parameters ($T_{\rm obs}$, $\rho$ and $r$) are ignored. Their contributions are minor, and the total uncertainty is dominated by that from the stellar properties, $\frac{\sigma_\rho}{3\rho_\star}$ as shown in Fig.~\ref{fig_errors}.  } 

In the single-transiting case, the only free model parameter is the mean eccentricity $\bar{e}$.   

\subsection{3.2 Multiple Transiting Systems}
For multiple transiting systems, the method is the same as above except for drawing inclination $\cos i_0$  distribution in the \emph{Step 2} of Section 3.1. 

{For single transiting systems,  the $\cos i_0$ from different systems are independent of each other. However, this is not the case for multiple transiting systems, where $\cos i_0$ of different planets in the same system are correlated.}
In order to take into account the above effect, for a multiple transiting system, we first draw a reference $\cos i^{'}_{0}$ uniformly, then for each transiting planet in the system, similarly to  Fabrycky et al.\cite{Fab14}, we set its $\cos i_0$ as
\begin{eqnarray}
\cos i_0 =\cos i^{'}_{0} + R_N*\sigma_i,
\label{cosi}
\end{eqnarray}
where $R_N$ is a random number drawn from a normal distribution centered at 0 and with standard deviation of 1, resulting in a Rayleigh distribution of width $\sigma_i$ in mutual inclination $i$, namely
\begin{eqnarray}
dN = \frac{i}{\sigma_{i}^2}{\rm exp}\left(-\frac{{i}^2}{2\sigma_{i}^2}\right)d i,
\label{dni}
\end{eqnarray} 
where $\sigma_i$ is Rayleigh parameter and the mean inclination $\bar{i}=\sigma_i\sqrt{\pi/2}$.

In the multiple transiting case, there are two free parameters for modeling the transit duration ratio, which are the mean eccentricity $\bar{e}$ and the mean inclination $\bar{i}$.  {\xie Note that we simulate transit duration ratios system by system. A simulated system is selected if all the simulated planets in the system have enough S/N to be detected.} 

\section{Transit Duration Ratio: Observations vs Simulations}
In this section, we describe using Maximum Likelihood (hereafter ML) method to estimate the planet eccentricity and/or inclination distributions by modeling the observed transit duration ratio distributions.

We calculate the likelihood ${\bf L}$ as a function of $\bar{e}$ and/or $\bar{i}$, and the best-fit model has the maximum ${\bf L}$. The likelihood function ${\bf L}$ is computed similarly to Hadden \& Lithwick\cite{HL14}. The likelihood that a given planet in our sample has observed transit duration ratio ${\rm TDR}_{\rm obs}$ is 
\begin{eqnarray}
L({\rm TDR}_{\rm obs} |\bar{e},\bar{i}) = \nonumber \\ \int P({\rm TDR} |\bar{e},\bar{i}) {\rm exp}[-({\rm TDR}-{\rm TDR}_{\rm obs})^2/2\sigma_{\rm TDR}^2] d{\rm TDR}.
\label{like}
\end{eqnarray}
where the first term on the right hand side, $P({\rm TDR} |\bar{e},\bar{i})$, is the probability that the transit duration ratio as determined by the theoretical model (equation \ref{tdrmod}), assuming the model parameters $\bar{e}$ and $\bar{i}$ are randomly drawn from the Rayleigh distribution (equations \ref{dne} and \ref{dni}). The second term, ${\rm exp}[-({\rm TDR}-{\rm TDR}_{\rm obs})^2/2\sigma_{\rm TDR}^2]$, is the probability that the model $\rm TDR$ generates the observed one $\rm TDR_{obs}$ given the noise distribution. Here $\sigma_{\rm TDR}$ is the 1-sigma uncertainty of ${\rm TDR}_{\rm obs}$ (equation \ref{tdrsig}). We compute the total likelihood ${\bf L}$  by multiplying together the likelihoods for all planets.

\subsection{4.1 Single Transiting Systems}
In this case, the only fitting parameter is the mean orbital eccentricity $\bar{e}$. We consider  a series of $\bar{e}$ from 0.0001 (essentially 0) to 0.6  with an interval of 0.002. For each $\bar{e}$, following section 3.1, we generate 300 simulated transit duration ratio ${\rm TDR}_{\rm mod}$ for each planet to calculate the probability $P({\rm TDR} |\bar{e})$ in equation \ref{like}. The total likelihood, ${\bf L}$ is plotted as a function of $\bar{e}$ in Fig.~1 in the main text. The  likelihood values are then smoothed as a function of $\bar{e}$ using a spline function to estimate the confidence intervals (1-$\sigma$: 68.3$\%$, 2-$\sigma$:95.4$\%$ and 3-$\sigma$: 99.7$\%$).

\subsection{4.2 Multiple Transiting Systems}
In this case, the fitting parameter is the mean orbital eccentricity $\bar{e}$ (equation \ref{dne}) and mean mutual orbital inclination $\bar{i}$. We consider a grid of $\bar{i}-\bar{e}$, where $\bar{i}$ from 0.00001 (essentially 0) to 0.2 ($\sim11.5^\circ$) with an interval of 0.002 and $\bar{e}$ is from 0.0001 (essentially 0) to 0.4 with an interval of 0.01. For each pair of $\bar{e}$ and $\bar{i}$, following section 3.2, we generate 100 modeled transit duration ratios ${\rm TDR}_{\rm mod}$ for each planet to calculate the probability predicted by the model, i.e.,  $P({\rm TDR} |\bar{e},\bar{i})$ in equation \ref{like}. ${\bf L}$ is shown as a contour map in the plane of $\bar{i}-\bar{e}$ in Fig.~2 in the main text, which gives the confidence intervals (1-$\sigma$: 68.3$\%$, 2-$\sigma$:95.4$\%$ and 3-$\sigma$: 99.7$\%$) of $\bar{i}$ and $\bar{e}$.

\section{Further Discussions}

\subsection{5.1 An Abrupt Transition or A Smooth Correlation?}
Recently, Limbach \& Turner \cite{LT15} analyzed the RV planet sample 
and they reported that the planetary eccentricity is anti-correlated to planetary multiplicity -- the mean eccentricity progressively decreases with the number of planets in the system.  
In our work, as shown in Fig.~\ref{fig_ecc_inc_np}, all $\bar{e}$ in the three multiple subsamples are comparable within uncertainties and close to zero, which is in contrast to the relatively large $\bar{e}$ in the single subsample ($N_p=1$). This suggests an abrupt transition rather than a smooth correlation 
as a function of number of transiting planets in the system. 
However, we caution that it is challenging to make a direct comparison between these two works: First, the majority of RV planets are Jovian planets, while most {\it Kepler} planets are super-Earths/Sub-Neptunes. Second, the detection efficiency and selection bias are different between RV and {\it Kepler} transit surveys, thus the number of planets in the system $N_p$ have different meanings between the two works. For example, some single transiting systems can come from intrinsically multiple-planet systems with only one planet transiting (see more discussions in the next section).

\subsection{5.2 Impact Parameter Distribution}
Transit light curves contain information of impact parameter (e.g.,  Seager \& Mallen-Ornelas \cite{SM03}) and in principle, such information can be incorporated into modeling the transit duration 
ratio distributions. In practice, inferences of individual impact parameters 
are most reliable for the {\it Kepler} planets with short-cadence (1 min) data, good knowledge of limb darkening and/or deep transits (i.e., high SNRs)  \cite{Pri15}. Previous works (e.g., see Fig.~9 of Swift et al. \cite{Swi15}) have shown that impact parameter is difficult to determine from the long
cadence data. As our sample is mainly composed of small planets (thus relatively shallow depth) with long cadence (30 minutes) data, we choose not to 
use the derived impact parameters for individual objects. Instead, we model the impact parameter distribution from simulation as given in supplementary Section 3. 

\begin{figure*}
\centering
\includegraphics*[width=\textwidth]{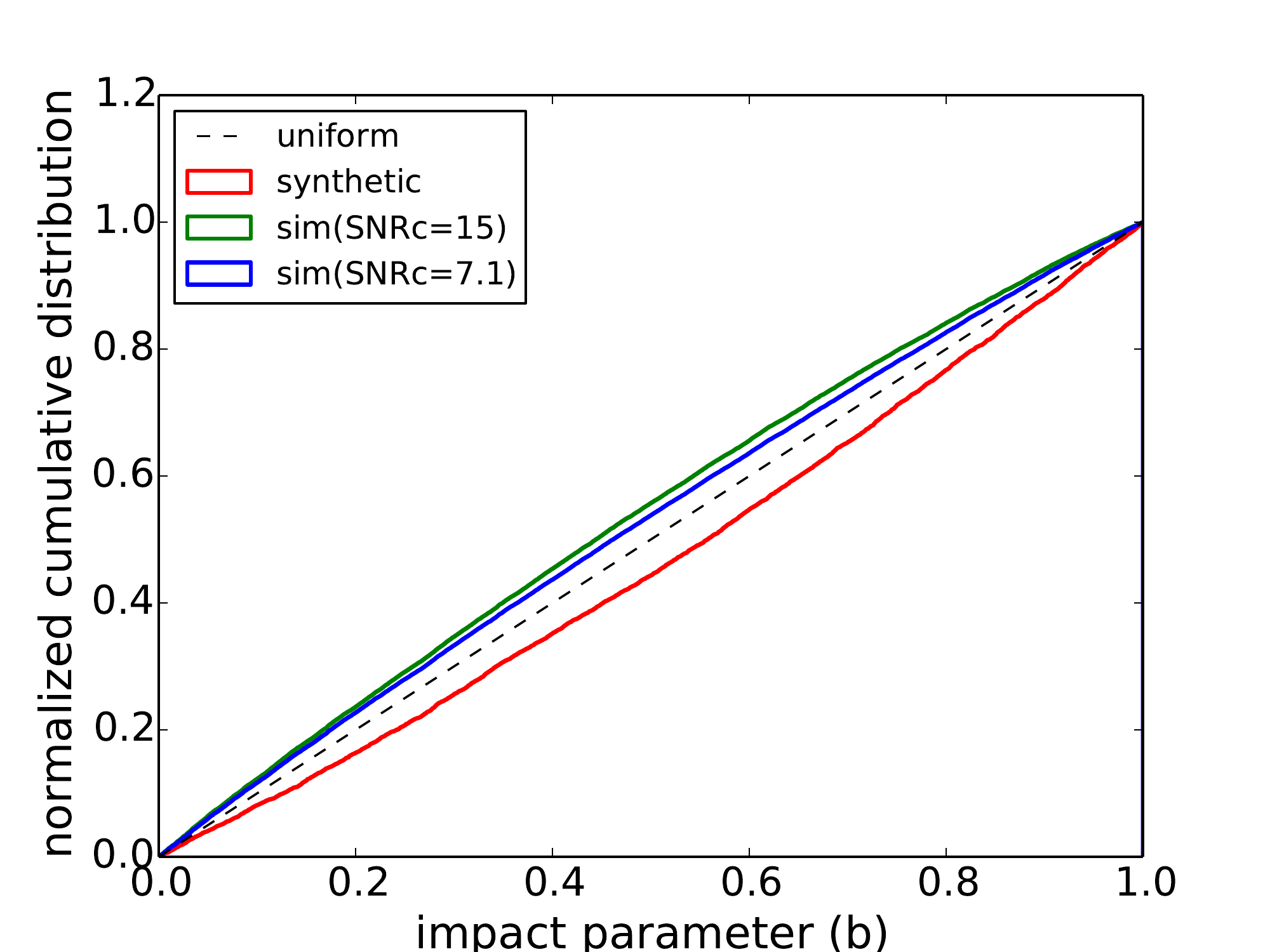}
\caption{{\bf Transit impact parameter distributions of single transiting systems.} The blue and green curves depict the distributions resulting from our nominal simulations with signal noise ratio cuts, SNRc=7.1 and SNRc=15 (see Supporting Information Sec. 3.1 and 5.5). The red curve is the one generated from a synthetic model discussed in Supporting Information Sec. 5.2), and the black dashed line shows the uniform distribution for comparison.}
\label{fig_single_check_imp}
\end{figure*}

In the following, we discuss the impact parameter distributions of the singles, which are plotted in Fig.~\ref{fig_single_check_imp}.  The blue and green lines depict the distributions in our nominal simulation with signal noise ratio cut ${\rm SNR}_c=7.1$ and simulation with ${\rm SNR}_c=15$ (see section 5.5), respectively. The impact parameters are smaller compared to the uniform distribution (black dashed line)\cite{Swi15}, and this is because smaller impact parameters lead to larger transit duration and thus higher SNR (i.e., more detectable).

There is another additional possible bias for having relatively larger contribution from larger impact parameters for singles. If a significant fraction of single-transiting systems come from coplanar multiple planet systems, their impact parameters should be biased towards large value to avoid seeing outer planets. In the following, we show that this bias is minor.

We generate a synthetic single transiting population by following the method described by Fang \& Margot\cite{FM12}. Specifically, we first generate $10^6$ planetary systems assuming that each systems have $N$ planets with mutual orbital inclination of $i$. Here $N$ is drawn from a bounded uniform distribution represented by a single parameter $\lambda$, and $i$ is drawn from Rayleigh distribution with a scaling parameter of $\sigma_i$. We adopt $\lambda=2.5$ (best fit of Fang \& Margot\cite{FM12}), and $\sigma_i=2^\circ$ (motivated by our results shown in the Fig.~2 of the main text). We then arbitrarily select a viewing angle and choose the single transiting systems.  The red line in Fig.~\ref{fig_single_check_imp} shows the impact parameter distribution of the synthetic single transiting population. As compared to the uniform distribution, it biases towards large impact parameter as expected. Adopting the synthetic impact parameter distribution to fit the transit duration ratio, we obtain $\bar{e}=0.275^{+0.029}_{-0.026}$, which is consistent with our nominal result shown in the Fig.~1 of the main text.\\

\subsection{5.3 Outermost of Multiples} 
\begin{figure*}
\centering
\includegraphics*[width=1.05\textwidth]{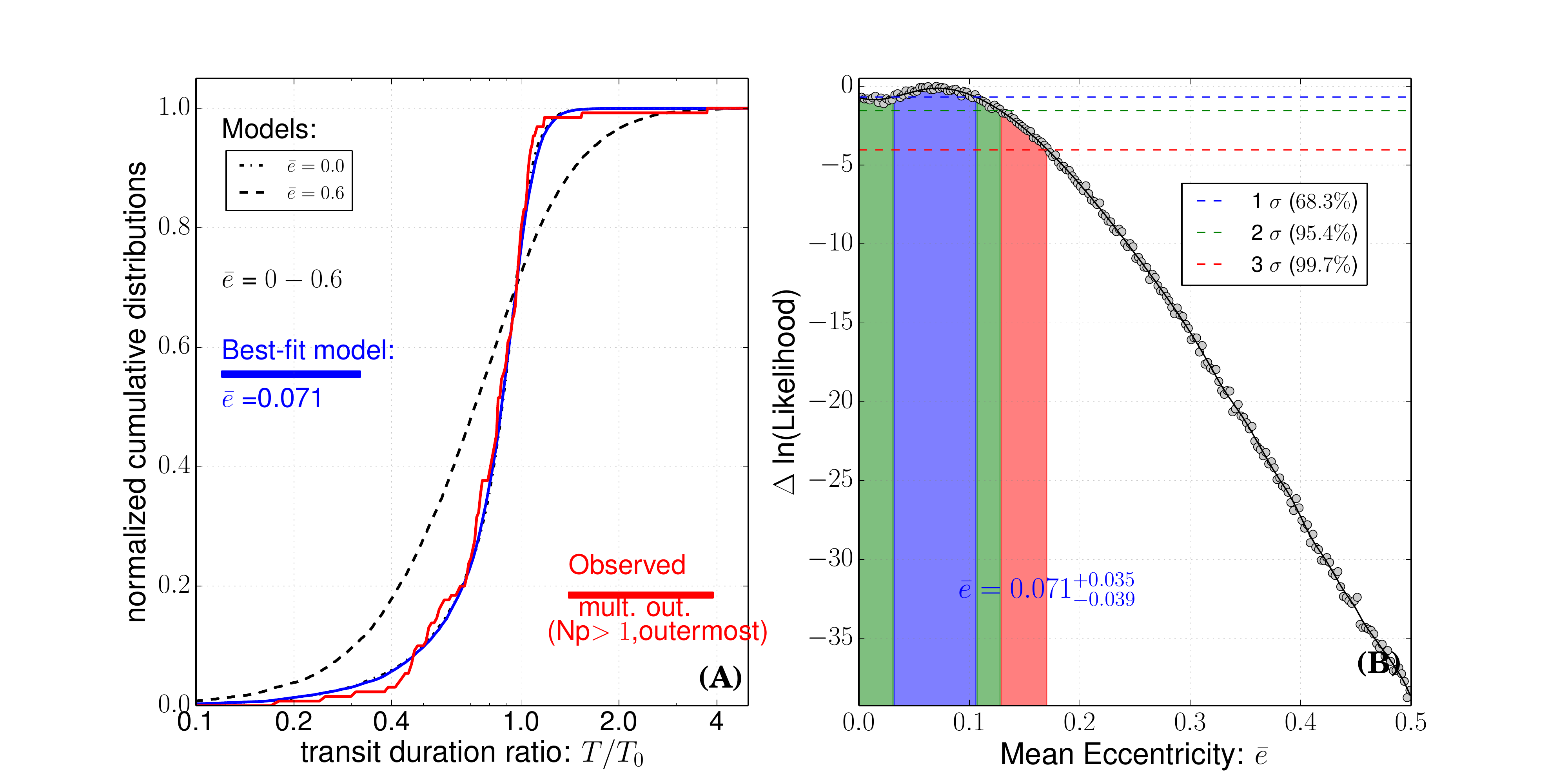}
\caption{Similar to Fig.~1 in the main text except showing the results for the outermost transits in multiple transiting systems. (See more discussions in Supporting Information Sec. 5.3).} 
\label{fig_ecc_mul_out}
\end{figure*}

For multiple transiting systems, if we only consider the outermost ones, then there is no need to fit the mutual inclination. In this case, we can do eccentricity-only fit as is done for the singles to perform a direct comparison. Fig.~\ref{fig_ecc_mul_out} shows the eccentricity fitting results for the outermost of multiples, which are consistent with nearly circular orbits in contrast to the relatively large eccentricities of singles (Fig.~1 in the main text). The results reinforce our conclusion:  on average, multiples are dynamically cold while singles are hot.

\subsection{5.4 False Positive}
{\dong 
The majority of {\it Kepler} planet candidates lack direct confirmation with RV, and it is important to assess how much the false positives (FPs) may affect the eccentricity distributions derived in our work. The overwhelming majority of {\it Kepler} multiples $(\sim 98\%)$ are believed to be {\it bona fide} planets \cite{Lis12}, thus the issue of FPs is most concerning for the single-transiting systems. We perform the following two sets of analyses to study the effects of FPs on our results.}

{\dong {\bf (1)} We attempt to eliminate KOIs with large estimated false positive probabilities (FPPs) by making various cuts on our sample.}

{\dong First, we remove subsets of our samples with large estimated FPPs. Indirect statistical estimates generally find a low $(\lesssim 10\%)$ averaged FPP for the entire sample of {\it Kepler} planet candidates \cite{MJ11, Fre13, Des15, Mor16} but the FPPs can be much higher for certain subsets of candidates. Approximately half of the {\it Kepler} giant planet candidates measured by Santern et al. \cite{San12, San15} with radial velocities are found to be FPs, and statistical studies also find that FPPs are substantially higher for large planet candidates (radius $>$6$R_\oplus$) \cite{Fre13, Mor16}. \cite{Col12} found that FPPs can also depend on orbital periods, and the close-in planet candidates with orbital period $<$3\,d may have higher FPPs. We remove the large ($>$6$R_\oplus$) and close-in ($<$ 3 d) planet candidates, yielding smaller samples with 280 and 291 planet candidates in the single and multiple transiting systems, respectively. The transit duration ratio distribution fits to these two samples give $\bar{e}=0.285^{+0.024}_{-0.023}$ for singles and $\bar{e}\le0.076$, $0.017<\bar{i}\le0.065$ for multipls, and they are consistent with the results shown in Fig.~1 and 2 in the main text.}

{\dong Second, we reduce the averaged FPP of our sample by eliminating the candidates
with large estimated FPPs by Morton et al.\cite{Mor16}. Recently Morton et al.\cite{Mor16} published their FPP estimates for all KOIs individually, and their estimates are consistent with existing direct RV measurements such as Santern et al. \cite{San15}.}

\begin{figure*}
\centering
\includegraphics*[width=0.8\textwidth]{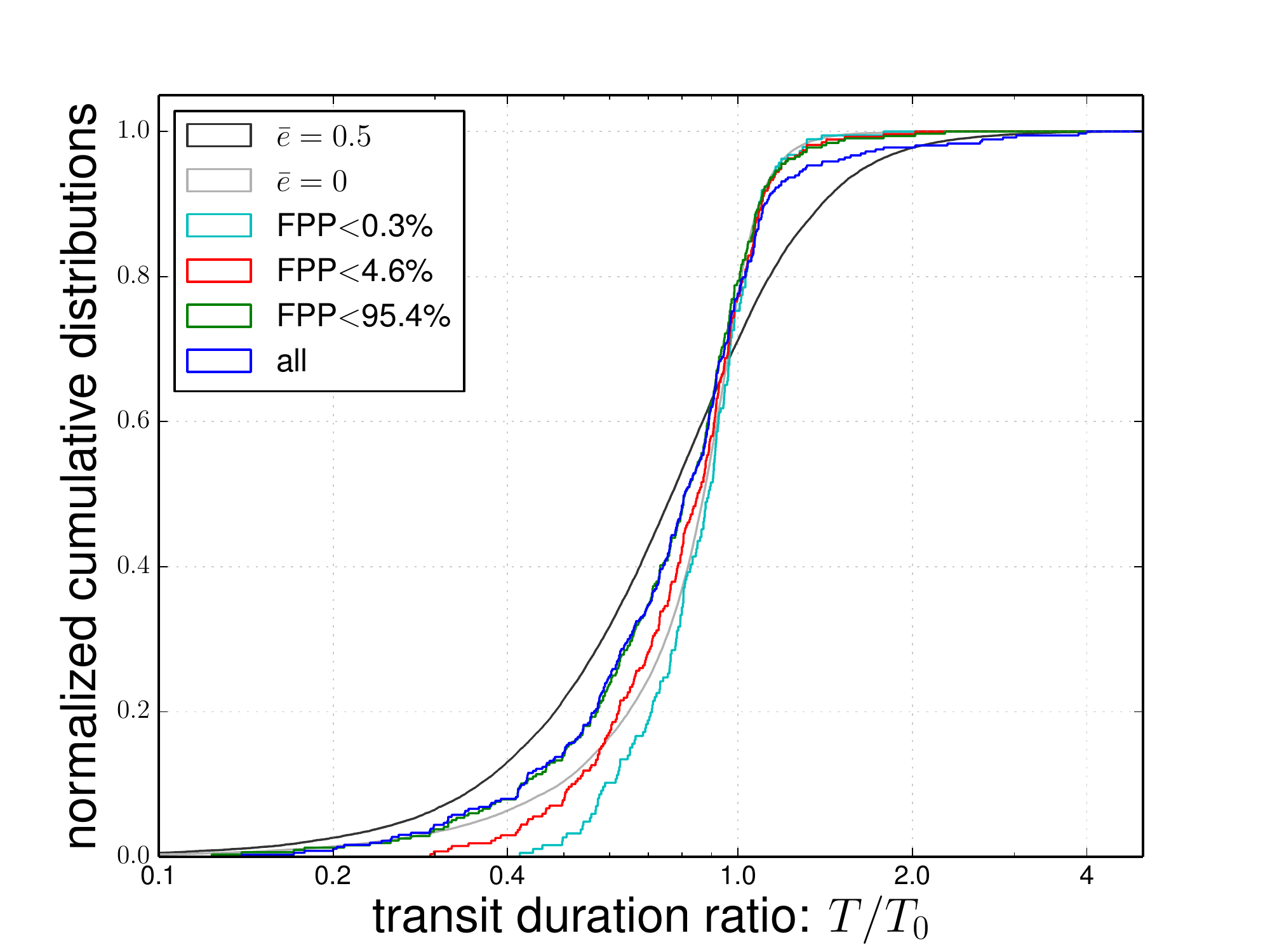}
\caption{Transit duration ratio distribution of single transiting KOIs with false positive probabilities (FPP) less than 95.4\%, 4.6\% and 0.3\%. For comparison, We also plot the distribution of all singles without the FPP cut and the modeling distributions with $\bar{e}=0$ and $\bar{e}=0.5$.} 
\label{fig_TDR_PC}
\end{figure*}

\begin{figure*}
\centering
\includegraphics*[width=0.8\textwidth]{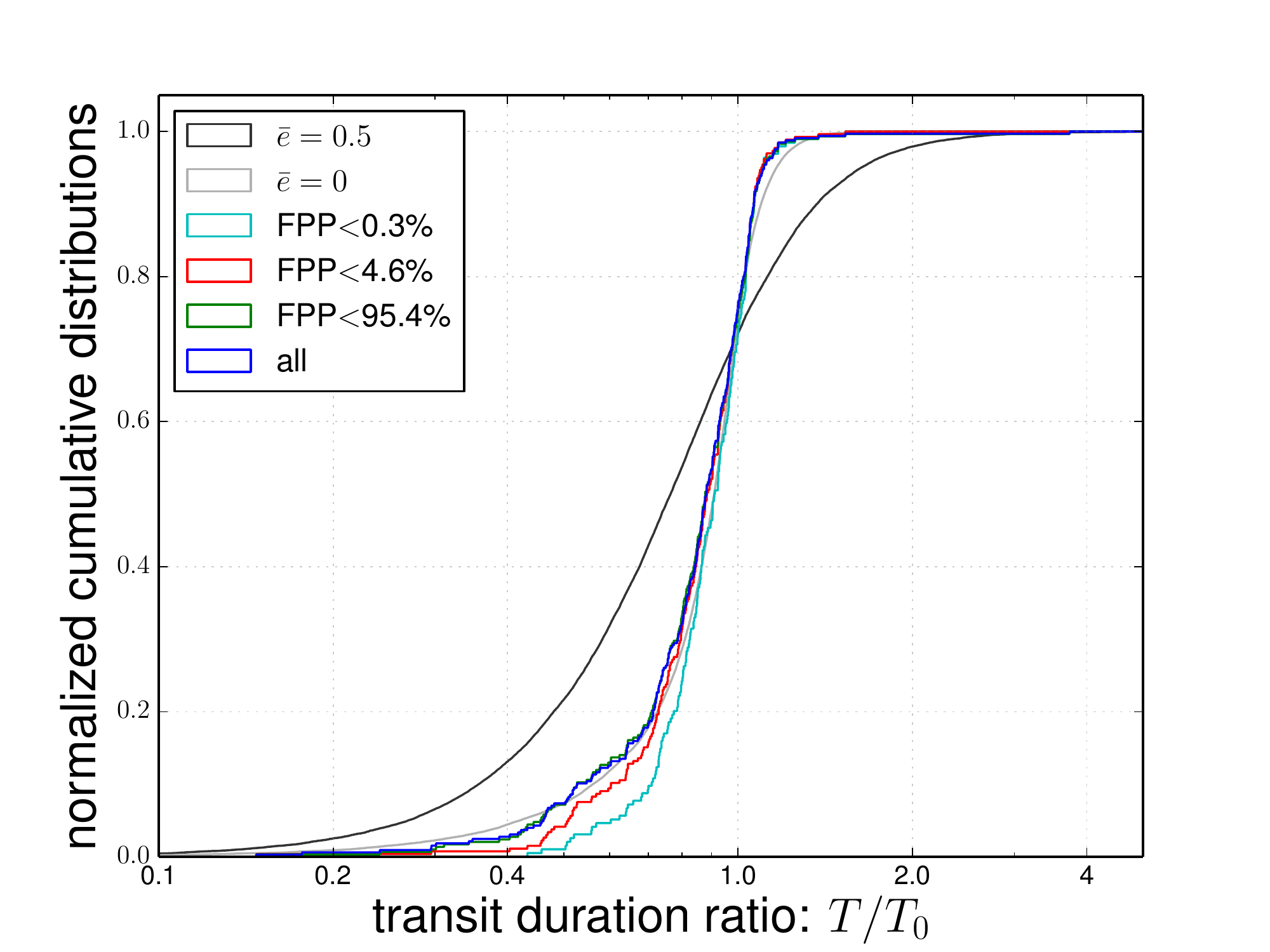}
\caption{Similar for Fig.\ref{fig_TDR_PC} but for multiple transiting KOIs.} 
\label{fig_TDR_PC_multi}
\end{figure*}

{\dong At first thought, one might proceed the analysis by selecting the KOIs with low FPP (e.g., FPP$<4.6\%$ and FPP$<0.3\%$, corresponding to 2-$\sigma$ and 3-$\sigma$ confidence for true planets). However, we find that only analyzing those with low FFPs can be problematic. Fig.\ref{fig_TDR_PC} shows the transit duration distributions of single-transiting planet candidates with various FPP cuts. The distributions for FPP$<4.6\%$ (2-$\sigma$; red) and FPP$<0.3\%$ (3-$\sigma$; cyan) differ significantly.
Both distributions are truncated at $T/T_0\gtrsim0.3-0.4$, while $T/T_0$ should get down to $\sim0$ (corresponding to impact parameter $b=1$) in any physically valid models. Similar 
patterns show up for multiple-transiting systems by performing the same 
FPP cuts (see Fig. \ref{fig_TDR_PC_multi}) and the resulting distributions are not 
consistent with any models (e.g., $\bar{e}=0-0.5$, gray and black lines), 
signifying the failure with this 
approach. A main problem with this approach is that, cutting at a low FPP threshold (e.g., FPP<4.6\%) excludes a significant fraction of KOIs with relatively high probabilities being true planets (e.g., FPP$\sim$10\% thus $\sim 90\%$ probability being true planets). While in Morton et al.\cite{Mor16}, transit duration is used as part of input information to infer FPP, so FPP may have some dependence on transit duration. For example, some true planets at high impact parameters and thus small transit duration ratios can have relatively large FPPs. Cutting the sample at low FPP thresholds can therefore remove true planets in a way that depend on their transit duration ratios, which in turn introduces a bias in the resulting 
transit duration ratio distribution. In our single-planet sample, according to 
Morton et al.\cite{Mor16}, the mean FPP is $\sim 12\%$, but performing 
cuts of FPP$<4.6\%$ and FPP$<0.3\%$ remove $\sim 27\%$ and $\sim 49\%$ of the 
sample, respectively. So the difference between the transit duration ratio distributions of the $<4.6\%$ and $<0.3\%$ FPP cuts are not mainly caused by eliminating FPs but rather by removing a large fraction of true planets in a biased fashion from the sample.}

\begin{figure*}
\centering
\includegraphics*[width=1.0\textwidth]{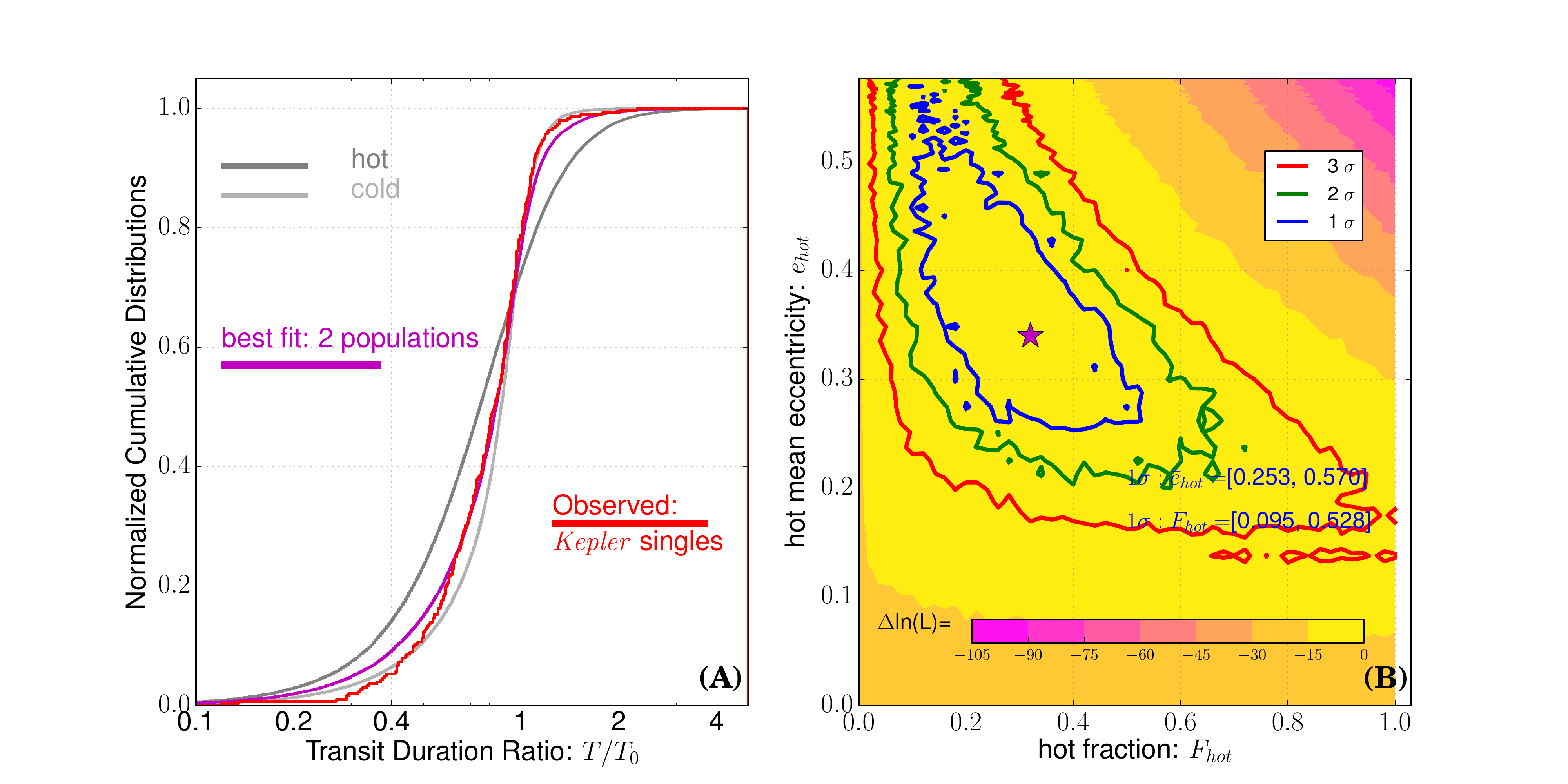}
\caption{Similar to Fig.4, but here we cut off observed KOIs with large false positive probability (FPP$>$68.3\%). For comparison, we also plot the transit duration ratio distributions of a hot ($\bar{e}$=0.6) and cold ($\bar{e}$=0.04) populations in the left panel.} 
\label{fig_single_bimodal_FPoff1}
\end{figure*}

\begin{figure*}
\centering
\includegraphics*[width=1.0\textwidth]{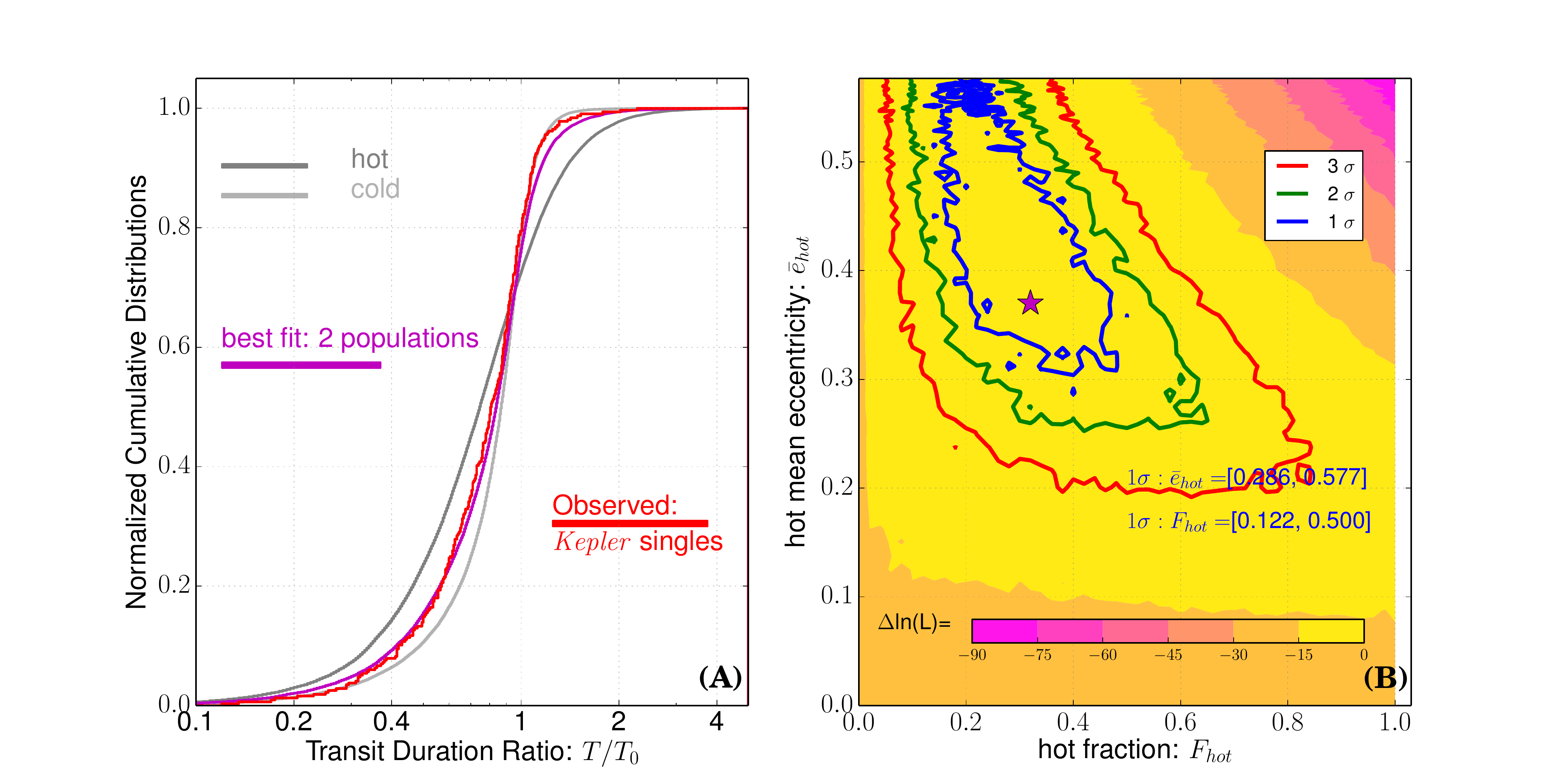}
\caption{Similar to Fig.4, but here we cut off observed KOIs with large false positive probability (FPP$>$95.4\%). For comparison, we also plot the transit duration ratio distributions of a hot ($\bar{e}$=0.6) and cold ($\bar{e}$=0.04) populations in the left panel.} 
\label{fig_single_bimodal_FPoff2}
\end{figure*}

{\dong Instead of keeping planet candidates with low FPPs, we choose to 
remove planet candidates with high FPPs. We find that in Morton et al. \cite{Mor16}, the FPPs of the sample are dominated by those with high FPPs. If we remove planet candidates with high FPPs (those with FPP>68.3\% and FPP>95.4\%), the mean FPP of the sample reduce to $\sim$2.2\% and $\sim$5.9\% respectively, which are comparable to the mean FPP ($\sim$2.4\% and $\sim$3.9\%) of multiple transiting systems by making the same FPP cuts. With such low mean FPPs, the effects of FPs should be nearly negligible. In Fig.\ref{fig_single_bimodal_FPoff1} and Fig.\ref{fig_single_bimodal_FPoff2}, we plot the results of two-population fit for singles using the two high FPP cuts (by keeping candidates with FPP<68.3\% and FPP<95.4\%) respectively. Qualitatively, the results are comparable to those in the main text without FPP cut (Fig.4), namely, they all reveal a hot and a cold populations in the singles. Quantitatively, the results are consistent with each other within their 1-$\sigma$ uncertainties. As compared to Fig.4, the best-fit mean eccentricities for the hot population are lower ($\sim 0.3 - 0.4$ as compared to $\sim 0.6$), while the fraction of the hot population are somewhat higher ($\sim 0.3$ as compared to $\sim 0.2$).}

\begin{figure*}
\centering
\includegraphics*[width=0.8\textwidth]{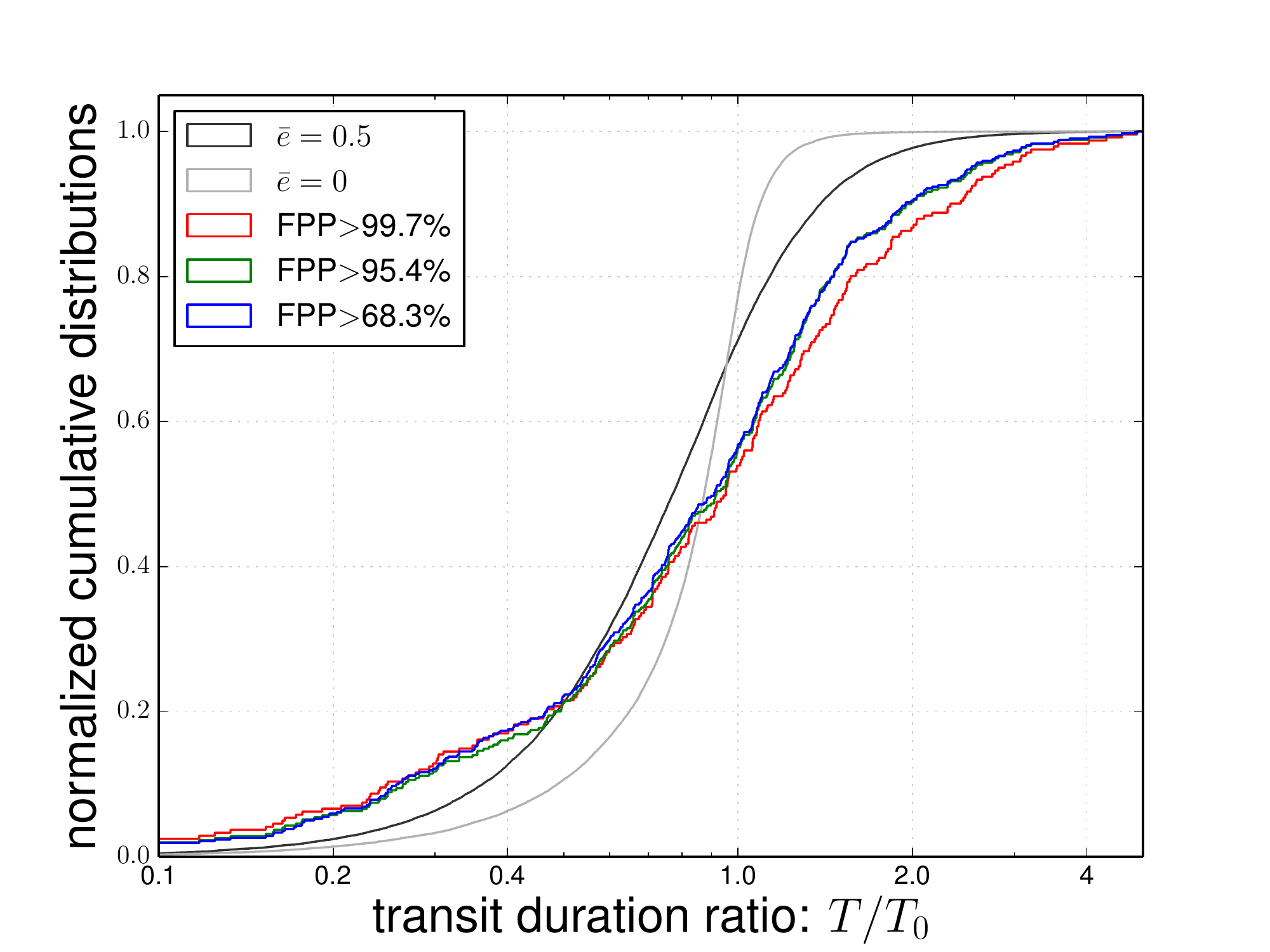}
\caption{Similar to Fig.\ref{fig_TDR_PC} but we focus on KOIs which are likely false positive with FPP greater than 68\% (1-sigma), 95\%(2-sigma) and 99.7\%(3-sigma).  } 
\label{fig_TDR_FP}
\end{figure*}

\begin{figure*}
\centering
\includegraphics*[width=1.0\textwidth]{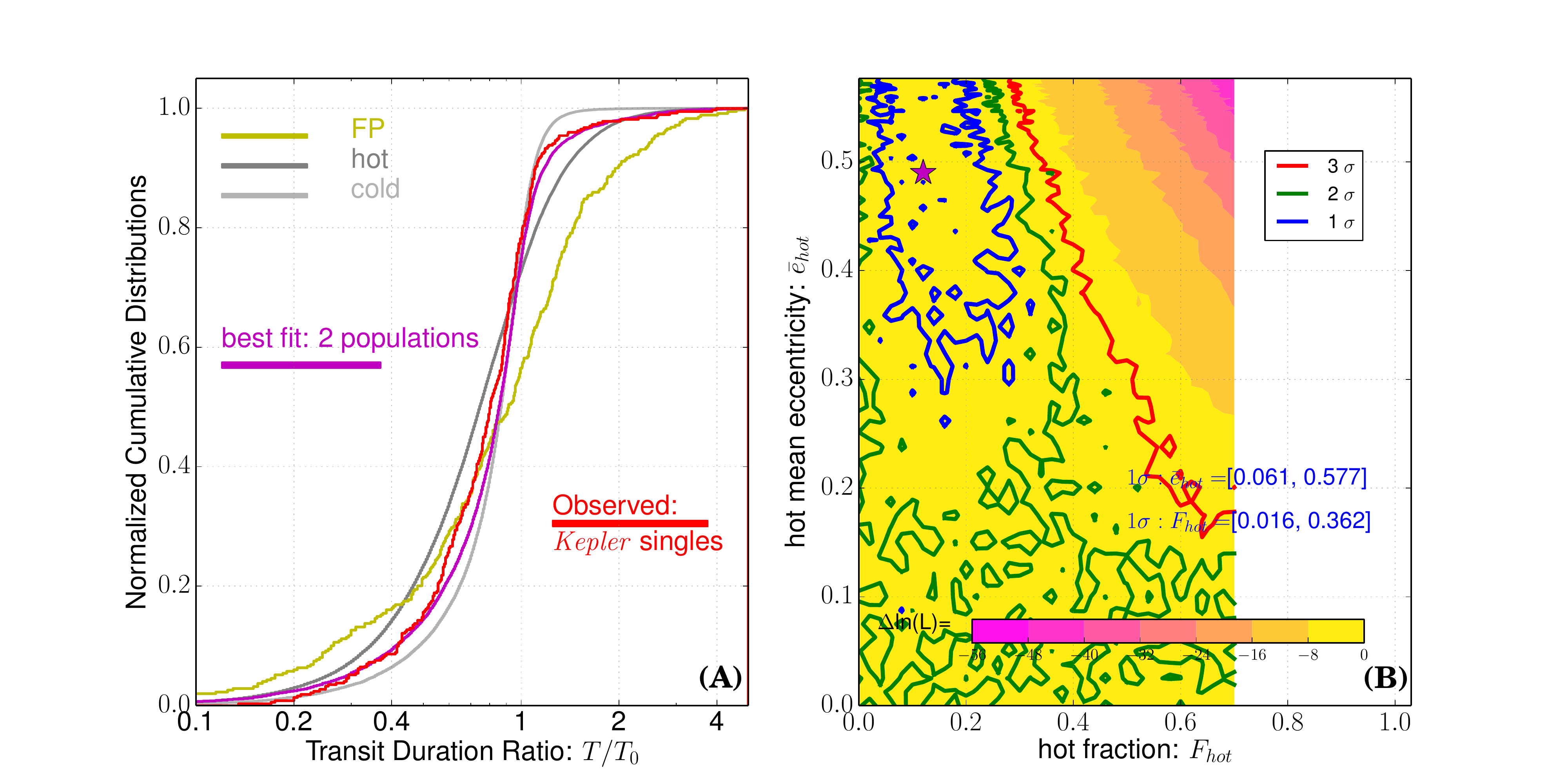}
\caption{Similar to Fig.4, but here we add in our model a fix component (12\% in fraction) with transit duration ratio distribution drawn from that of false positives (FPP>95.4\%).  For comparison, we also plot the transit duration ratio distributions of the false positive, the hot ($\bar{e}$=0.6) and cold ($\bar{e}$=0.04) populations in the left panel. In the right panel, the contour map is truncated because the hot fraction varies in a smaller range between 0 and 0.7 in our simulation. } 
\label{fig_single_bimodal_FPcut2}
\end{figure*}

\begin{figure*}
\centering
\includegraphics*[width=1.0\textwidth]{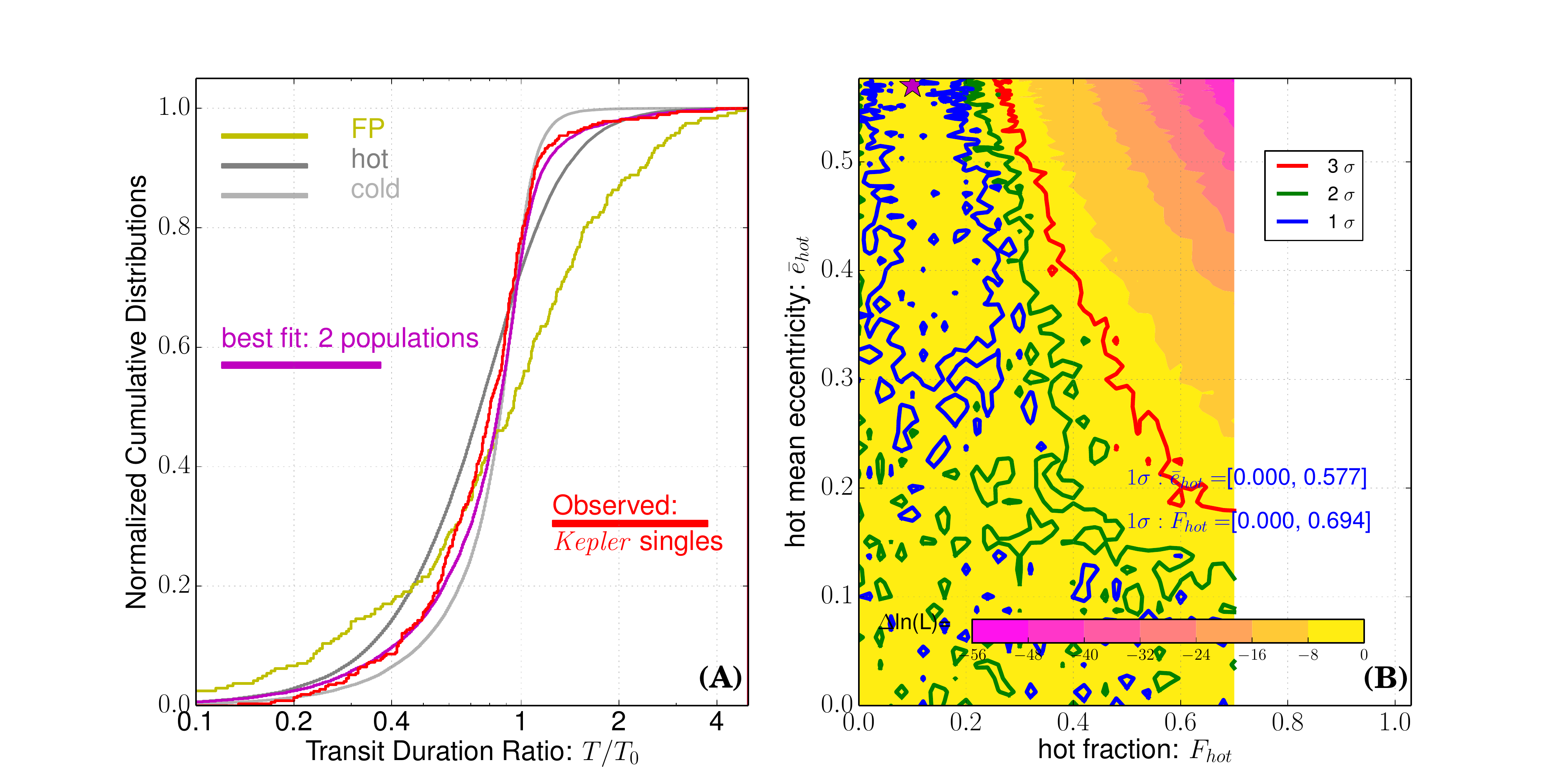}
\caption{Similar to Fig.\ref{fig_single_bimodal_FPcut2}, but here we add in our model a fix component (12\% in fraction) with transit duration ratio distribution drawn from that of false positives (FPP>99.7\%).} 
\label{fig_single_bimodal_FPcut3}
\end{figure*}

{\dong {\bf (2)} Alternatively, we try to model the impact of FPs by injecting FPs into our simulations. It is beyond the scope of our work to directly simulate the expected transit duration ratio distribution of FPs from first principles. We instead take an empirical approach by using the KOIs with large FPP according to Morton et al.\cite{Mor16}. In Fig.\ref{fig_TDR_FP}, we plot the transit duration ratio distributions of KOIs with FFP greater than 68.3\%, 95.4\% and 99.7\%, corresponding to 1-$\sigma$, 2-$\sigma$ and 3-$\sigma$ confidence levels of FPs. Their distributions (blue, green and red lines) are shallower than expected from the planet models (gray and black lines for $\bar{e}=0-0.5$). This is consistent with our qualitative expectation that the duration distribution should be wider for the transits/eclipses from the FPs. Furthermore, we see that the duration distribution of FP is not very sensitive to the FPP criteria. All three FP criteria lead to similar duration distributions while the 3-$\sigma$ one (red line in Fig.\ref{fig_TDR_FP}) has a somewhat larger deviation from the others between $T/T_0=1$ and $T/T_0=3$. Following this approach, we are not able to 
completely eliminate the true planets to obtain the ``pristine'' FP transit duration ratio distribution, and also there are systematic uncertainties due to the dependency of FPP estimates on transit duration. Despite these limitations, since the three distributions are similar to each other and consistent qualitatively with expectations from FPs, they may be useful in informing us the effects of FPs on the transit duration ratio distribution. In order to account for the systematic uncertainties as much as possible, below we use all three distributions presented in Fig.\ref{fig_TDR_FP} to model the FP distributions.

Using the FPP estimates given by Morton et al.\cite{Mor16}, we find that the mean FPP of singles in our original single-planet sample is $\sim$12\%. Thus, we inject 12\% simulated candidates with transit duration ratio $T/T_0$ drawn from the distributions of FPs by adopting various FP criteria as shown in Fig.\ref{fig_TDR_FP} and discussed in the previous paragraph. We then repeat the two-population fit as done in Fig.4. In Fig.\ref{fig_single_bimodal_FPcut2} and Fig.\ref{fig_single_bimodal_FPcut3}, we plot the results of using the 2-$\sigma$ (green line in Fig.\ref{fig_TDR_FP}) and the 3-$\sigma$ (red line in Fig.\ref{fig_TDR_FP}) FP criteria, respectively. The result of the 1-$\sigma$ criterion is nearly identical to the one using the 2-$\sigma$ criterion and is not shown. As can be seen in Fig.\ref{fig_single_bimodal_FPcut2} and Fig.\ref{fig_single_bimodal_FPcut3}, the results are consistent with those shown in Fig.4 of the main article -- the singles are composed of a major dynamically cold population and a minor dynamically hot population. However, we note that, the goodness of fit with modeling FP injection (Fig.\ref{fig_single_bimodal_FPcut2} or Fig.\ref{fig_single_bimodal_FPcut3}) is considerably worse than that of fit with performing FP cut from the sample ( Fig.\ref{fig_single_bimodal_FPoff1} and Fig.\ref{fig_single_bimodal_FPoff2}). This is not surprising -- in the FP injection approach, we implicitly assume that those ``hidden'' FPs with relatively 
low FPPs in the sample also follow the injected FP duration distributions selected
from high FPPs, but this assumption is almost certainly not warranted.

Based on the above two kinds of analyses, we therefore conclude that the main conclusions shown in the main text (Fig.4) is qualitatively sound despite the uncertainties introduced by FPs.}

\subsection{5.5 Signal-to-Noise Ratio}
In this work, the Signal-to-Noise Ratio for detection is cut at $SNR_c=7.1$ (Step 4 of Section 2.1) for both simulation and observation. To see how the SNR cut affects our results, we vary the threshold of SNR, i.e., ${\rm SNR}_c=10$ and 15,  and repeat the same analyses. We find the results are all {\dong consistent with} those shown in the Fig.~1 and 2 in the main text, suggesting that the adopted SNR cut is unlikely to affect our results. 

\subsection{5.6 Stellar Property Calibrator}
As mentioned in Section 1, we derived two sets of uncertainties of stellar properties (e.g. $log(g)$) based on the calibrators of high-resolution spectroscopy and seismology respectively. All the above results are based on the seismology calibrator. {For comparison, we have performed the same analyses as {\dong those resulting} in Fig.~1 and 2 but using the stellar properties derived from the  calibrator of high-resolution spectroscopy. We find that both calibrators generally give {\dong consistent} results, i.e., the singles are dynamically hot with {\dong mean} eccentricities about 0.2-0.3, while the multiples are dynamically cold with eccentricities close to zero. Nevertheless, we note that the {\dong mean} eccentricity of the singles derived from the spectroscopy calibrator is $0.21\pm0.04$, which is somewhat smaller (by about 2 $\sigma$) than that derived from the seismology calibrator. }

\newpage

\clearpage

\end{document}